# Branching ratios, radiative lifetimes and transition dipole moments for YbOH

Ephriem Tadesse Mengesha, Anh T. Le and Timothy C. Steimle*

*School of Molecular Science*

*Arizona State University*

*Tempe, Arizona 85287*

Lan Cheng and Chaoqun Zhang

*Department of Chemistry*

*The Johns Hopkins University*

*Baltimore, Maryland, 21218 U.S.A.*

Benjamin L. Augenbraun, Zack Lasner and John Doyle

*Physics Department*

*Harvard University*

*Cambridge, Massachusetts 02138*

*corresponding: tsteimle@asu.edu

**Abstract**

Medium resolution ($\Delta\tilde{\nu} \sim 3$ GHz) laser-induced fluorescence (LIF) excitation spectra of a rotationally cold sample of YbOH in the 17300-17950 cm$^{-1}$ range have been recorded using two-dimensional (excitation and dispersed fluorescence) spectroscopy. High resolution ($\Delta\lambda \sim 0.65$ nm) dispersed laser induced fluorescence (DLIF) spectra and radiative decay curves of numerous bands detected in the medium resolution LIF excitation spectra were recorded. The vibronic energy levels of the $\tilde{X}\,^2\Sigma^+$ state were predicted using a discrete variable representation approach and compared with observations. The radiative decay curves were analyzed to produce fluorescence lifetimes. DLIF spectra resulting from high resolution ($\Delta\tilde{\nu} < 10$ MHz) LIF excitation of individual low-rotational lines in the $\tilde{A}\,^2\Pi_{1/2}(0,0,0)-\tilde{X}\,^2\Sigma^+(0,0,0)$, $\tilde{A}\,^2\Pi_{1/2}(1,0,0)-\tilde{X}\,^2\Sigma^+(0,0,0)$, and $[17.73]\Omega=0.5(0,0,0)-\tilde{X}\,^2\Sigma^+(0,0,0)$ bands were also recorded. The DLIF spectra were analyzed to determine branching ratios which were combined with radiative lifetimes to obtain transition dipole moments. The implications for laser cooling and trapping of YbOH are discussed.

## I. INTRODUCTION

Laser-cooled, linear triatomic molecules (e.g., YbOH) may be a sensitive venue for investigating *T*-violating physics beyond the Standard Model (BSM).[1] This BSM physics includes the determination of the electron electric dipole moment (eEDM) and/or the nuclear magnetic quadrupole moment (MQM). Both eEDM and MQM result in a parity- and time-reversal-violating molecular EDM, although originating from different underlying physics.[2] The effective electric field ($E^{\mathrm{eff}}$) and the time-reversal-violating magnetic quadrupole moment interaction constant ($W_{\mathrm{M}}$) relevant to these probes of BSM physics in YbOH are predicted to be similar to that of YbF[3–5], which has been used previously for eEDM measurements[6]. The use of YbOH has experimental advantages due to the presence of parity doublets arising from the metastable, degenerate bending modes. The $^{171}$YbOH isotopologue has also been considered as a venue for detecting nuclear spin dependent symmetry-violating effects.[7] The precision of these proposed measurements would be greatly improved by laser cooling of YbOH to micro-kelvin temperatures and subsequent trapping, thus providing long coherence times.

Cooling and trapping seems possible given the success of laser cooling of SrF[8], CaF[9], YbF[10] and YO[11]. Like these molecules, YbOH has electronic transitions involving metal-centered, non-bonding electrons which are very diagonal (i.e. vibration selection rule of $\Delta v = 0$). In comparison to SrF, CaF, YO and YbF, the requisite spectroscopic data needed for the design and implementation of YbOH laser cooling/trapping is rather limited. The high temperature, Doppler limited, spectroscopic study of various bands of the $\tilde{A}^2\Pi_{1/2} - \tilde{X}^2\Sigma^+$ and $\tilde{A}^2\Pi_{3/2} - \tilde{X}^2\Sigma^+$ electronic transitions was reported some time ago.[12] We have reported on the rotational spectroscopy[13] of $^{174}$YbOH and, more recently, on the analysis of the high-resolution electronic spectroscopy[14] of the $\tilde{A}^2\Pi_{1/2}(0,0,0)-\tilde{X}^2\Sigma^+(0,0,0)$, $\tilde{A}^2\Pi_{1/2}(0,0,0)-\tilde{X}^2\Sigma^+(1,0,0)$ and $\tilde{A}^2\Pi_{1/2}(1,0,0)-\tilde{X}^2\Sigma^+(0,0,0)$ bands of supersonic molecular beam samples of $^{172}$YbOH and $^{174}$YbOH. The latter study included optical Stark and Zeeman spectroscopy for the $\tilde{A}^2\Pi_{1/2}(0,0,0)-\tilde{X}^2\Sigma^+(0,0,0)$ band of $^{174}$YbOH which were analyzed to determine the magnitude of the molecular frame permanent electric dipole moment, $|\vec{\mu}_{el}|$, and magnetic g-factors for the $\tilde{X}^2\Sigma^+(0,0,0)$ and $\tilde{A}^2\Pi_{1/2}(0,0,0)$ states. An additional band near 17730 cm$^{-1}$, designated as the [17.73]0.5- $\tilde{X}^2\Sigma^+(0,0,0)$ transition, has also been recorded and analyzed at high spectral resolution[15]. It is noteworthy that the [17.73]0.5 state has a significantly shorter inter-nuclear separation than that of the $\tilde{A}^2\Pi_{1/2}$ and $\tilde{X}^2\Sigma^+$ states, suggestive of an $\left([\mathrm{Xe}]4f^{13}\right)_{\mathrm{Yb}^+}\sigma^2_{\mathrm{Yb}^+(6s6p)}$ dominant configuration similar to the [561] and [571] states of YbF[16].

The scheme for cooling and trapping YbOH[1] involves direct laser slowing of the forward velocity of molecules emanating from a cryogenic buffer gas source[17,18] to a degree sufficient for three-dimensional magneto-optical trapping and transfer to a conservative optical dipole trap. Very recently, YbOH was laser-cooled in one dimension[19] and a subset of the data reported here was instrumental for that measurement. In that study the transverse temperature of the YbOH beam was reduced by nearly two orders of magnitude to less than 600 μK and the phase space density increased by a factor of greater than six via Sisyphus cooling. To extend these results to allow for three-dimensional cooling and trapping will require detailed knowledge of vibrational branching ratios from a variety of electronic states in order to achieve a large number of photons to be scattered (>10,000) per molecule with minimal loss to dark vibrational sublevels.

In this work we report spectroscopic measurements necessary for implementation of efficient photon cycling and determination of repumping schemes required for laser cooling and trapping as well as supporting vibronic energy level calculations. The data includes experimentally measured fluorescence branching ratios, $b_{iv',fv''}$, and fluorescence lifetimes, $\tau_{\mathrm{iv}'}$, which are used to obtain transition dipole moments (TDMs) associated with the numerous visible transitions of $^{174}$YbOH. Analogous to the laser cooling scheme for YbF[5], the main cooling transitions for $^{174}$YbOH will be the $^{P}P_{11}(1)$ ($\tilde{\nu}$ =17323.5699 cm$^{-1}$) and $^{P}Q_{12}(1)$ ($\tilde{\nu}$ =17323. 5952 cm$^{-1}$) lines[14] of the $\tilde{A}\,^2\Pi_{1/2}(0,0,0)-\tilde{X}\,^2\Sigma^+(0,0,0)$ origin band. The $^{P}P_{11}(1)$ and satellite $^{P}Q_{12}(1)$ lines are rotationally closed excitations. Vibrational repumping will be required to recover population that decays to levels other than the $\tilde{X}\,^2\Sigma^+(0,0,0)$ state. Repumping pathways can involve either direct $\tilde{X}\,^2\Sigma^+(v_1,v_2,v_3)\rightarrow\tilde{A}\,^2\Pi_{1/2}(0,0,0)$ excitation or indirect pumping into an excited state other than $\tilde{A}\,^2\Pi_{1/2}(0,0,0)$ which spontaneously emits to $\tilde{X}\,^2\Sigma^+(0,0,0)$:

$$\tilde{X}\,^2\Sigma^+(v_1'',v_2'',v_3'')\overset{laser}{\longrightarrow}\text{excited}(v_1',v_2',v_3')\overset{spon.}{\longrightarrow}\tilde{X}\,^2\Sigma^+(0,0,0)\quad. \tag{1}$$

Repumping of both the Yb–O stretching, $\nu_1$, and the Yb–O–H bending, $\nu_2$, vibrational levels of the $\tilde{X}\,^2\Sigma^+$ state will be required for efficient laser cooling. Excited O-H stretching modes, $\nu_3$, should not strongly coupled to either the $\tilde{X}\,^2\Sigma^+(0,0,0)$ or $\tilde{A}\,^2\Pi_{1/2}(0,0,0)$ levels.

In the present dispersed fluorescence measurements, the rotational fine structure is not resolved and the determined property is the rotationally averaged Einstein $A$-coefficient, $A_{\mathrm{iv}',fv''}$:

$$A_{\mathrm{iv}',fv''}=b_{iv',fv''}\tau_{\mathrm{iv}'}^{-1}=\frac{64\pi^4}{(4\pi\varepsilon_0)3h}\left|\mu_{iv',fv''}\right|^2\nu_{iv',fv''}^3=3.137\times10^{-7}\left|\mu_{iv',fv''}\right|^2\nu_{iv',fv''}^3, \tag{2}$$

where $\mu_{iv',fv''}$ is the vibronic TDM and $h$ is Planck's constant. The numerical coefficient of Eq. 2 assumes units of Debye (D) and wavenumbers (cm$^{-1}$). The $\mu_{iv',fv''}$ values are useful for determining the Einstein $A$-coefficient for rotationally resolved transitions, $A_{v'J',v''J''}$, relevant for optical pumping of individual quantum levels:

$$A_{v'J',v''J''} = 3.137\times10^{-7}\times\frac{\tilde{v}^3_{v'J',v''J''}S_{v'J',v''J''}}{(2J'+1)} = 3.137\times10^{-7}\times\frac{\tilde{v}^3_{v'J',v''J''}S_{J',J''}\left(\mu_{iv',fv''}\right)^2}{(2J'+1)} \quad . (3)$$

In Eq. 3, $S_{v'J',v''J''}$ is the line strength factor and $S_{J',J''}$ the Hönl-London factor, which is readily obtained from the eigenvalues and eigenvectors for the ground and excited state.

## II. EXPERIMENTAL

A rotationally cold ($T^{rot.}$ ~20 K) YbOH sample was generated by laser ablation of a continuously rotating ytterbium metal rod in a supersonic expansion of room temperature vapor of 50% hydrogen peroxide solution seeded in argon at a stagnation pressure of ~2 MPa. The production of YbOH was approximately a factor of four larger when methanol is substituted for the hydrogen peroxide solution, but in that case the spectra are contaminated with $YbOCH_3$. Initial detection was achieved by performing pulsed dye laser survey scans using a two-dimensional (2D) (excitation and dispersed laser induced fluorescence (DLIF)) spectroscopic technique[20,21] in the 17300 cm$^{-1}$ and 17950 cm$^{-1}$ spectral range. Typically, the 2D spectra were obtained by co-adding the signal of 20 laser ablation samples at each pulsed dye laser wavelength. The entrance slit width of the monochromator was set to 2 mm resulting in an approximately ± 4 nm DLIF spectral resolution, and a 1 μs detection window used for the intensified charge coupled detector (ICCD) attached to the monochromator. Subsequently, higher resolution DLIF spectra resulting from either medium-resolution pulsed dye laser or single frequency cw-dye laser excitation were recorded. For these measurements, the monochromator slits were typically reduced approximately 0.2 mm resulting in an approximately ± 0.4 nm spectral resolution and 10,000 laser ablation samples were averaged. DLIF measurements of rotationally resolved excitations were also performed. In these measurements low-$J$ branch features of the $\tilde{A}^2\Pi_{1/2}(0,0,0)-\tilde{X}^2\Sigma^+(0,0,0)$, $\tilde{A}^2\Pi_{1/2}(1,0,0)-\tilde{X}^2\Sigma^+(0,0,0)$ and [17.73]0.5-$\tilde{X}^2\Sigma^+(0,0,0)$ bands were excited using cw-dye laser radiation. The relative sensitivity of the spectrometer as a function of wavelength was calibrated using a black-body radiation source. The wavelength calibration of the DLIF spectra was achieved by recording the emission of an argon pen lamp. Fluorescence lifetime measurements were performed by tuning the wavelength of the pulsed dye laser to intense bandheads and monitoring the DLIF spectrum with a relatively wide (1 μs) ICCD detection window. The detection window was progressively stepped further in time from the incident pulsed laser in 10 or 20 nanosecond increments. The resulting fluorescence decay curves were fit to a first order exponential to determine the upper state fluorescence lifetimes $\tau_{\mathrm{i}v'}$.

## III. COMPUTATIONS

Electronic ground state vibrational levels of YbOH were obtained from a four-dimensional discrete variable representation (DVR) calculation. (For a review of DVR methods, see Ref.[22]) The present DVR calculations used grid points in reduced normal coordinate representation and employed an established DVR formulation for rectilinear coordinates[23]. The normal coordinates were obtained from harmonic frequency calculations using equation-of-motion electron-attachment singles and doubles (EOMEA-CCSD) method[24] and correlation-consistent core-valence triple-zeta basis set for Yb[25] and cc-pVTZ basis sets for O and H[26] . Scalar-relativistic effects were taken into account using the spin-free exact two-component theory in its one-electron variant (SFX2C-1e)[27,28]. Electronic energies on 2178 grid points covering the local potential energy surface up to around 10000 cm$^{-1}$ above the energy of the equilibrium structure were computed using the SFX2C-1e-EOMEA-CCSD method and subsequently fit into a six-order polynomial function of three internal coordinates (Yb-O bond length, O-H bond length, and Yb-O-H angle). The analytic potential function thus obtained reproduces *ab initio* energies reasonably well, with a maximum deviation of 6.1 $cm^{-1}$ and a root mean square deviation of 1.2 $cm^{-1}$. Details about the potential energy surface and the normal coordinates are enclosed in the Supporting Information. In the DVR calculations, potential energies on the DVR grid points in normal coordinate representation were obtained on the fly by referencing to the analytic potential function in internal coordinates. We used 21 evenly spaced DVR points covering the range [-4.0q, 4.0q] for Yb-O stretching mode, 28 points covering the range [-4.0q, 6.8q] for O-H stretching mode, and 21 points covering the range [-4.0q, 4.0q] for each bending mode, where q denotes corresponding dimensionless reduced normal coordinate. Computed vibrational levels were converged to below 2 $cm^{-1}$ with respect to the range and density of DVR grid points used here. Note that we used a seemingly excessive range for O-H stretching mode, which is needed to sample the correct region for Yb-O-H bending when combined with displacements of bending modes. It is perhaps more efficient to use more sophisticated curvilinear-coordinate representations for DVR calculations of polyatomic molecules[29–31]. Nevertheless, the DVR calculations presented here with simple normal-coordinate representation were affordable. Besides, the use of normal coordinates is naturally consistent with an important motivation of the present study to obtain accurate splitting between ($02^00$) and ($02^20$) vibronic levels of the $\tilde{X}\,^2\Sigma^+$ state. All calculations were carried out using the CFOUR program package[32–35] [11-14].

## IV. OBSERVATIONS

The transition wavenumbers, associated energy levels and assignments for the eleven most intense bands observed in the 17300 $cm^{-1}$ to 17950 $cm^{-1}$ spectral range are presented in Figure 1.

The lower energy levels associated with these transitions are all assigned to various $(v_1, v_2^l, v_3)$ vibrational levels in the $\tilde{X}\,^2\Sigma^+$ state. The numbers next to the $\tilde{X}\,^2\Sigma^+$ $(v_1, v_2^l, v_3)$ quantum numbers in Figure 1 are the observed energies in wavenumbers. The bands at 17323, 17730 and 17908 cm$^{-1}$ have been recorded and analyzed at high spectral resolution and are assigned to the $\tilde{A}\,^2\Pi_{1/2}(0,0,0) - \tilde{X}\,^2\Sigma^+(0,0,0)$, $[17.73]0.5 - \tilde{X}\,^2\Sigma^+(0,0,0)$ and $\tilde{A}\,^2\Pi_{1/2}(1,0,0) - \tilde{X}\,^2\Sigma^+(0,0,0)$ transitions, respectively. The origins, taken as the frequency of the $^{Q}Q_{11}(0)$ line, of these bands are precisely known[14,15] to be 17323.6500, 17730.5874 and 17907.8571 cm$^{-1}$. The band at 17378 cm$^{-1}$ is readily assigned to $\tilde{A}\,^2\Pi_{1/2}(1,0,0) - \tilde{X}\,^2\Sigma^+(1,0,0)$ transitions based upon combination differences of the known origins of the $\tilde{A}\,^2\Pi_{1/2}(0,0,0) - \tilde{X}\,^2\Sigma^+(0,0,0)$, $\tilde{A}\,^2\Pi_{1/2}(0,0,0) - \tilde{X}\,^2\Sigma^+(1,0,0)$ and $\tilde{A}\,^2\Pi_{1/2}(1,0,0) - \tilde{X}\,^2\Sigma^+(0,0,0)$ bands. The assignment of the remaining seven bands at 17332, 17345, 17637, 17643, 17681, 17708, and 17900 cm$^{-1}$ is more speculative. The previous study[12] of a high temperature sample assigned a band near 17339 cm$^{-1}$ to the $\tilde{A}\,^2\Pi_{1/2}(0,1^1,0) - \tilde{X}\,^2\Sigma^+(0,1^1,0)$ transition and observed bands at 17638 ± 5, 17681 ± 5, and 17729 ± 5 cm$^{-1}$ to $\tilde{A}\,^2\Pi_{1/2}(0,1^1,0) - \tilde{X}\,^2\Sigma^+(0,0,0)$ transitions. The observed DLIF spectra and radiative lifetimes (see below) of the 17345 and 17681 cm$^{-1}$ bands suggests that these bands have a common excited state. The 336 cm$^{-1}$ separation of these two bands corresponds to the energy of the $\tilde{X}\,^2\Sigma^+(0,1^1,0)$ state. The common excited state will simply be labelled as [17.68] although it has some $\tilde{A}\,^2\Pi_{1/2}(0,1^1,0)$ character. Based upon the DLIF spectra (see below) the five remaining bands at 17332, 17637, 17643, 17708 and 17900 cm$^{-1}$ are assigned as excitation from the $\tilde{X}\,^2\Sigma^+(0,0,0)$ state to excited states of unknown character which are labelled as [17.33], [17.637], [17.643], [17.71] and [17.90], respectively.

The 2D spectrum in the 17315 to 17385 $cm^{-1}$ region, which is near the origin band, is presented in Figure 2. The majority of the fluorescence occurs at the laser excitation wavelength ('on resonance') as expected for the strong bands of YbOH, which are promotions of non-bonding, metal-centered electrons. There is weak emission to the blue (anti-Stokes shifted) by approximately -11 nm (≈330 $cm^{-1}$) in the 17340 to 17355 $cm^{-1}$ range and stronger emission to the red (Stokes shifted) by approximately +18 nm (≈530 $cm^{-1}$) for laser excitation at 17323, 17332 and 17375 $cm^{-1}$. Three excitation spectra extracted by vertical integration of the signal along the horizontal slices indicated in Figure 2 are presented in Figure 3: on resonance ("Ex1"), Stokes shifted by one quantum of Yb-OH stretching excitation $\nu_1''$ ("Ex2"), and anti-Stokes shifted by one quantum of bending excitation, $\nu_2''$ ("Ex3"). The on-resonance and Stokes shifted excitation spectra exhibit blue degraded bands at 17323 $cm^{-1}$, 17332 $cm^{-1}$ and 17375 $cm^{-1}$. A broad, weak, band in the 17345-17355 $cm^{-1}$ range is evident in the anti-Stokes shifted spectrum. Also presented in Figure 3 is the predicted LIF excitation spectrum obtained using the derived spectroscopic parameters[14] for the $\tilde{X}\,^2\Sigma^+(0,0,0)$, $\tilde{X}\,^2\Sigma^+(1,0,0)$, $\tilde{A}\,^2\Pi_{1/2}(0,0,0)$ and $\tilde{A}\,^2\Pi_{1/2}(1,0,0)$ states.

High-resolution DLIF spectra recorded by tuning the pulsed dye laser to 17323, 17332, 17345 and 17375 $cm^{-1}$ are presented in Figure 4, along with the associated energy levels. The numbers above the spectral features are the measured shifts in wavenumber ($cm^{-1}$) relative to the laser. Excitation of the $\tilde{A}\,^2\Pi_{1/2}(0,0,0)-\tilde{X}\,^2\Sigma^+(0,0,0)$ band at 17323 $cm^{-1}$ results in emission to the $\tilde{X}\,^2\Sigma^+(0,0,0)$ and $\tilde{X}\,^2\Sigma^+(1,0,0)$ levels while excitation of the $\tilde{A}\,^2\Pi_{1/2}(1,0,0)-\tilde{X}\,^2\Sigma^+(1,0,0)$ band at 17375 $cm^{-1}$ results in emission to the $\tilde{X}\,^2\Sigma^+(0,0,0)$, $\tilde{X}\,^2\Sigma^+(1,0,0)$, $\tilde{X}\,^2\Sigma^+(0,2^0,0)$ and $\tilde{X}\,^2\Sigma^+(2,0,0)$ levels. The excitation of the weak band near 17345 $cm^{-1}$, which is most evident in the excitation spectrum extracted from the horizontal slice of the 2D spectrum shifted to the blue (anti-Stokes ) from the laser by one quantum of the $\tilde{X}\,^2\Sigma^+$ bending (~330 cm$^{-1}$) (Figure 3), produced emission to the primarily to $\tilde{X}\,^2\Sigma^+(0,1^1,0)$ but also weakly to the $\tilde{X}\,^2\Sigma^+(0,0,0)$, $\tilde{X}\,^2\Sigma^+(1,0,0)$ $\tilde{X}\,^2\Sigma^+(1,1^1,0)$ and $\tilde{X}\,^2\Sigma^+(2,0,0)$ levels. The appearance of vibronically induced $\Delta\nu_2 = \pm1$ features in this DLIF spectrum illustrates that the [17.68] state is not purely the $\tilde{A}\,^2\Pi_{1/2}(0,1^1,0)$ state. The relatively strong, blue degraded band near 17332 cm$^{-1}$ (Figure 2), which was not reported in the previous study[12], is unassigned. The DLIF spectrum resulting from excitation of the 17332 cm$^{-1}$ band is very nearly identical to that resulting from excitation to the $\tilde{A}\,^2\Pi_{1/2}(0,0,0)$ state (Figure 4). This band is assigned as the $[17.33]-\tilde{X}\,^2\Sigma^+(0,0,0)$ transition based upon the observed lack of anti-Stokes spectral features. Although the intensities of the 17332 cm$^{-1}$ and 17323 cm$^{-1}$ bands are comparable under the relatively high pulsed laser intensity used to record the 2D spectrum of Figure 2, under lower pulsed laser intensities the $\tilde{A}\,^2\Pi_{1/2}(0,0,0)-\tilde{X}\,^2\Sigma^+(0,0,0)$ band at 17323 cm$^{-1}$ is approximately 5 times more intense than the $[17.33]-\tilde{X}\,^2\Sigma^+(0,0,0)$ band at 17332 cm$^{-1}$.

The 2D spectrum in the 17630 to 17750 cm$^{-1}$ spectral range is presented in Figure 5. The bands near 17681, 17708 and 17730 cm$^{-1}$ exhibit relatively strong and narrow emission shifted by multiple quanta of Yb-OH stretching $\nu_1''$ (≈+18 nm ≈530 cm$^{-1}$). The 17681 cm$^{-1}$ band also exhibits emission shifted by one quantum of Yb-OH bending $\nu_2''$ (≈+11 nm ≈330 cm$^{-1}$). Also evident in the 2D spectrum is a weak, sharp band at 17637 cm$^{-1}$ that has very diagonal fluorescence and a weak, spectrally broad band in the 17640 to 17660 cm$^{-1}$ range. The weak, non-horizontal, emission in the 2D spectrum blue shifted from the laser by approximately -11.5 nm (≈365 cm$^{-1}$) at 17630 nm is an artifact of amplified stimulated emission (ASE) of the pulsed dye laser. The ASE is exciting the very strong $6s^2\ ^1S_0 \rightarrow 6s6p\ ^3P_1$ transition of Yb(I) at 17992.007 cm$^{-1}$ (555.8nm) giving rise to an emission that is shifted from the laser wavelength (17630-17992≈ 365 cm$^{-1}$).

The medium resolution excitation spectra extracted from the horizontal slices of the 2D spectrum of Figure 5 taken on-resonance ("Ex1"), Stokes shifted by one quanta of $\nu_2''$, ("Ex2"), and Stokes shifted by one quanta of $\nu_1''$ ("Ex3") are presented Figure 6. The strong band at 17730 cm$^{-1}$, designated as the [17.73]0.5-$\tilde{X}\,^2\Sigma^+(0,0,0)$ transition, has been recorded and analyzed at high spectral resolution[15]. The much weaker 17637, 17681, and 17708 cm$^{-1}$ bands are all sharp and blue degraded, whereas the very weak 17643 cm$^{-1}$ band, which is most evident in the $\nu_2''$ Stokes shifted excitation spectrum, is broader and unstructured.

The high-resolution DLIF spectra resulting from pulsed laser excitation of the bandheads at 17637, 17643, 17681, 17708, and 17730 cm$^{-1}$ are presented in Figure 7 along with an associated energy level diagram. The numbers above the spectral features are the measured shifts in wavenumber (cm$^{-1}$) relative to the laser. The DLIF spectra for the weak 17637 and 17643 bands were recorded at slightly lower resolution to enhance the signal to noise ratio. The DLIF spectra from the [17.73]0.5 and [17.71] states exhibit long progression in the $\nu_1$ stretching mode and weaker emission shifted by intervals of $\nu_1+2\nu_2$. The [17.68] state emits to $\nu_1$ levels (i.e. $\tilde{X}\,^2\Sigma^+(1,0,0)$ and $\tilde{X}\,^2\Sigma^+(2,0,0)$), the $\tilde{X}\,^2\Sigma^+(0,1^1,0)$ level, and weakly to the $\tilde{X}\,^2\Sigma^+(1,1^1,0)$ level. The [17.643] state emits with high efficiency to the $\tilde{X}\,^2\Sigma^+(0,1^1,0)$ level and weakly to the $\tilde{X}\,^2\Sigma^+(0,0,0)$ and $\tilde{X}\,^2\Sigma^+(1,1^1,0)$ levels. The [17.637] state emits with high efficiency to the $\tilde{X}\,^2\Sigma^+(0,0,0)$ level and weakly to $\tilde{X}\,^2\Sigma^+(0,1^1,0)$ and $\tilde{X}\,^2\Sigma^+(1,0,0)$ levels.

The 2D spectrum in the 17880 to 17920 cm$^{-1}$ range is presented in the bottom portion of Figure 8. The excitation spectrum recorded by monitoring the emission Stokes shifted by one quantum of $\nu_1''$ (≈+18 nm ≈530 cm$^{-1}$) is presented at the top of Figure 8. The band at 17908 cm$^{-1}$ is the $\tilde{A}\,^2\Pi_{1/2}(1,0,0)-\tilde{X}\,^2\Sigma^+(0,0,0)$ transition which has been recorded and analyzed at high spectral resolution[14]. The nature of the excited state associated with the 17900 cm$^{-1}$ band is unknown. Although the intensities of the 17900 cm$^{-1}$ and 17908 cm$^{-1}$ bands are comparable under the relatively high pulsed laser intensity used to record the 2D spectrum of Figure 8, under lower pulsed laser intensities the $\tilde{A}\,^2\Pi_{1/2}(1,0,0)-\tilde{X}\,^2\Sigma^+(0,0,0)$ transition at 17908 cm$^{-1}$ is approximately three times more intense than the $[17.90]-\tilde{X}\,^2\Sigma^+(0,0,0)$ band at 17900 cm$^{-1}$. The DLIF spectra resulting from pulsed dye laser excitation at 17900 and 17908 cm$^{-1}$ bandheads are presented in Figure 9 along with the associated energy levels and assignments. The DLIF spectra resulting from excitation of the $\tilde{A}\,^2\Pi_{1/2}(1,0,0)$ and [17.90] states are very similar and dominated by emission to the $\tilde{X}\,^2\Sigma^+(1,0,0)$ state.

Fluorescent decay data for the $\tilde{A}\,^2\Pi_{1/2}(0,0,0)$, [17.33], [17.90] and $\tilde{A}\,^2\Pi_{1/2}(1,0,0)$ levels are presented in Figure 10, while those for the [17.637], [17.643], [17.68] and [17.73]0.5 states are presented in Figure 11. The DLIF signal from the [17.71] state was too weak to obtain a reliable decay curve. Also presented are the curves obtained by fitting the data points to an exponential decay. The $\tilde{A}\,^2\Pi_{1/2}(0,0,0)$, [17.33], [17.643], [17.90] and $\tilde{A}\,^2\Pi_{1/2}(1,0,0)$ levels all have relatively short lifetimes ranging from between 20 ± 2 ns for the $\tilde{A}\,^2\Pi_{1/2}(0,0,0)$ level to 35 ± 6 ns for the [17.643] level. The lifetimes for the [17.637], [17.68] and [17.73]0.5 states are significantly longer.

Decay curves obtained by exciting the $[17.68]-\tilde{A}^2\Pi_{1/2}(0,1^1,0)$ band at 17345 $cm^{-1}$ and monitoring both the on-resonance and anti-Stokes emission shifted by one quantum of Yb-OH bending $\nu_2''$ (≈-11 nm ≈330 $cm^{-1}$) are presented in Figure 12. The 17345 $cm^{-1}$ band is overlapped with the more intense $\tilde{A}^2\Pi_{1/2}(0,0,0)-\tilde{X}^2\Sigma^+(0,0,0)$ and $[17.33]-\tilde{X}^2\Sigma^+(0,0,0)$ bands (see Figure 3). The anti-Stokes emission is primarily that from the [17.68] level while the on-resonance emission is dominated by that due to excitation of the overlapping $\tilde{A}^2\Pi_{1/2}(0,0,0)-\tilde{X}^2\Sigma^+(0,0,0)$ and $[17.33]-\tilde{X}^2\Sigma^+(0,0,0)$ bands. The determined relatively long lifetime (110±15 ns) for the anti-Stokes emission is consistent with that determined for the emission resulting from excitation of the $[17.68]-\tilde{X}^2\Sigma^+(0,0,0)$ transition at 17680 $cm^{-1}$ (89±8ns). The on-resonance emission exhibits a short lifetime (21±4 ns) similar to that of the $\tilde{A}^2\Pi_{1/2}(0,0,0)$ and [17.33] states which are simultaneously excited at 17345 $cm^{-1}$.

The DLIF spectra resulting from excitation of the ${}^{P}P_{11}(1)$ lines of the $\tilde{A}^2\Pi_{1/2}(0,0,0)-\tilde{X}^2\Sigma^+(0,0,0)$ ($\tilde{\nu}$ =17323.5699 $cm^{-1}$), $\tilde{A}^2\Pi_{1/2}(1,0,0)-\tilde{X}^2\Sigma^+(0,0,0)$ ($\tilde{\nu}$ =17907.9028 $cm^{-1}$) and $[17.73]0.5-\tilde{X}^2\Sigma^+(0,0,0)$ ($\tilde{\nu}$ =17731.9707 $cm^{-1}$), which are used in the optical pumping scheme[19], are presented in Figure 13 and compared with pulsed dye laser DLIF spectra recorded at similar resolution. The sensitivities of the cw-dye laser measurements are less than those of the pulsed dye laser measurements. Hence, the entrance slit width of the monochromator was relatively large and the spectra of Figure 13 are broader than those of Figures 4, 7 and 9.

## V. ANALYSIS

The fluorescence decay curves were fit to a first-order exponential to determine the upper state fluorescence lifetimes $\tau_{i\nu'}$ which are presented in Table 1. The predicted decay curves using the fitted lifetimes are also presented in Figures 10, 11 and 12. The DLIF spectra resulting from excitation of the $\tilde{A}^2\Pi_{1/2}(0,0,0)$, [17.33], [17.637], [17.643], [17.68], [17.73]0.5, [17.90] and $\tilde{A}^2\Pi_{1/2}(1,0,0)$ states were corrected for wavelength sensitivity and the integrated peak areas used to determine the branching ratios, $b_{iv',fv''}$, which are also presented in Table 1. The $[17.68]-\tilde{X}^2\Sigma^+(0,0,0)$ transition at 17681 cm$^{-1}$ was used, as opposed to $[17.68]-\tilde{X}^2\Sigma^+(0,1^1,0)$ transition at 17345 cm$^{-1}$, because it is more intense and not overlapped. Also presented in Table 1 are the magnitudes of the transition dipole moments, $\left|\mu_{iv',fv''}\right|$, obtained using Eq. 2 and the measured $v_{iv',fv''}$, $\tau_{iv'}$ and $b_{iv',fv''}$ values.

## VI. DISCUSSION

### A. Vibrational levels of the $\tilde{X}^2\Sigma^+$ state.

The spacing between $\tilde{X}^2\Sigma^+(0,0,0)$ and $\tilde{X}^2\Sigma^+(1,0,0)$ has been previously determined[14] to be 529.3269(3) cm$^{-1}$. The observation of multiple vibronically induced bands in the excitation and DLIF spectra in the present study made it possible to determine the energies of eight additional vibrational levels of the $\tilde{X}^2\Sigma^+$ state up to 2100 cm$^{-1}$. The energies of the eight $\tilde{X}^2\Sigma^+$ state vibrational levels as determined from the DLIF spectra are listed in Table 2 and have estimated errors on the order of ± 5 cm$^{-1}$ depending upon signal-to-noise ratio and proximity to argon emission calibration lines. As noted above, there is strong experimental evidence that the bands at 17345 cm$^{-1}$ and 17681 cm$^{-1}$ share a common upper level. Specifically, the lifetime determined from the anti-Stokes emission of the 17345 cm$^{-1}$ band (Figure 11) is similar to that of the 17681 cm$^{-1}$ band as are DLIF spectral patterns (Figures 4 and 7). Note that although the patterns are similar, the relative intensities of the two DLIF differ, because the 17345 band is overlapped with the more intense $\tilde{A}^2\Pi_{1/2}(0,0,0)-\tilde{X}^2\Sigma^+(0,0,0)$ and $[17.33]-\tilde{X}^2\Sigma^+(0,0,0)$ bands. Under the assumption that these two bands have a common excited state, as indicated in Figure 1, the energy of the $\tilde{X}\,^2\Sigma^+(0,1^1,0)$ level is 336 ± 5 cm$^{-1}$ which is consistent with the value of 329 ± 5 cm$^{-1}$ obtained from the DLIF spectrum.

Also presented in Table 2 are the *ab initio* vibrational energies obtained from discrete variable representation (DVR) calculations. The *ab initio* energy levels agree fairly well with measured ones, with a maximum deviation of 10 cm$^{-1}$ for the (1,2$^2$,0) state. A predicted splitting of 24 cm$^{-1}$ between (02$^0$0) and (02$^2$0) states seems reasonable, compared with the corresponding value of 30 cm$^{-1}$ in SrOH[36]. This indicates that scalar-relativistic equation-of-motion coupled-cluster technique used here is capable of providing accurate description of local potential energy surface for the ground state of YbOH, and the four-dimensional DVR approach appears to be an appropriate method for calculations of vibrational levels for linear triatomic molecules. The observed and *ab initio* predicted state vibrational energies were fit to the phenomenological expression[37]:

$$E(\tilde{X}\,^2\Sigma^+, v_1, v_2, l_2) = \sum_{i=1,2}\omega_i(v_i + \frac{d_i}{2}) + \sum_{i=1,2}\sum_{k=1,2} x_{ik}(v_i + \frac{d_i}{2})^2 + g_{22}l_2^2 - E(\tilde{X}\,^2\Sigma^+, 0,0,0), \quad (4)$$

where $d_i$ is the degeneracy of the vibrational mode (i.e. 1 and 2 for $\nu_1$ and $\nu_2$, respectively). The results are presented in Table 2. The determined $g_{22}$ parameter (~ 5 cm$^{-1}$) is consistent with 7.56 cm$^{-1}$ value determined for the $\tilde{X}\,^2\Sigma^+$ state of SrOH[38]. The levels in the 0-2100 cm$^{-1}$ energy range that have not been experimentally detected are $\tilde{X}\,^2\Sigma^+(0,2^2,0)$, $\tilde{X}\,^2\Sigma^+(1,2^2,0)$, $\tilde{X}\,^2\Sigma^+(0,3^1,0)$, $\tilde{X}\,^2\Sigma^+(0,3^3,0)$, $\tilde{X}\,^2\Sigma^+(1,3^1,0)$ and $\tilde{X}\,^2\Sigma^+(1,3^3,0)$.

## B. The excited states

Nine excited vibronic states have been identified in the relatively small (≈ 650 cm$^{-1}$) spectral region probed. Simple molecular orbital considerations provide some insight into the cause of this high density of states and suggests that there may be at least nine Hund's case (a) low-lying excited electronic states. In simplest terms, these nine states are the $^2\Pi_r$ and $^2\Sigma^+$ states arising from a $4f^{14}6p^1$ (Yb$^+$) configuration, the $^2\Pi_i$, $^2\Phi_i$, $^2\Delta_i$ and $^2\Sigma^+$ states from the $4f^{13}6s^2$ (Yb$^+$) configuration, and the $^2\Delta_r$, $^2\Pi_r$ and $^2\Sigma^+$ states from the $4f^{14}5d^1$ (Yb$^+$) configuration. The multiple vibronic levels of these nine electronic states will strongly interact due to large spin-orbit and vibronic coupling.

The state at 17323 cm$^{-1}$ is analogous to the $A^2\Pi_{1/2}(v=0)$ ($E$=18106 cm$^{-1}$) state of YbF. In the case of YbF there are two Ω=1/2 states, often labeled as [557] and [561], which are 474 and 593 cm$^{-1}$ above the $A^2\Pi_{1/2}(v=0)$ state. Similarly, for YbOH there are two Ω=1/2 states, the [17.73]0.5 and $\tilde{A}^2\Pi_{1/2}(1,0,0)$, observed at 407 and 592 cm$^{-1}$ above the $\tilde{A}^2\Pi_{1/2}(0,0,0)$ state. In the case of YbF, the branching ratios[39] to the $X^2\Sigma^+$ ($v$=0, 1 and 2) levels from the [557] state are 13.2 , 70.7 and 13.9%, which are similar to those from the [561] state of 2.8, 89.0 and 7.8% , respectively. The fine and hyperfine parameters for the [557] and [561] states are also similar, which has been used to suggest that these two states are strong admixtures of an $A^2\Pi_{1/2}(v=1)$ state of the $4f^{14}6p^1$ (Yb$^+$) configuration and a perturbing state having an $\left([\mathrm{Xe}]4f^{13}\right)_{\mathrm{Yb}^+}\sigma^2_{\mathrm{Yb}^+(6s6p)}$ dominant configuration. Unlike YbF, the YbOH branching ratios (Table 1) for the two Ω=1/2 states at 407 and 592 cm$^{-1}$ above the $\tilde{A}^2\Pi_{1/2}(0,0,0)$ state are very dissimilar with the emission from the [17.73]0.5 state being predominantly to the $\tilde{X}^2\Sigma^+(0,0,0)$ level and that for the $\tilde{A}\,^2\Pi_{1/2}(1,0,0)$ being predominately to the $\tilde{X}\,^2\Sigma^+(1,0,0)$ level. The fine structure parameters[14,15] and lifetimes (Table 1) for the [17.73]0.5 and $\tilde{A}\,^2\Pi_{1/2}(1,0,0)$ states are also very dissimilar, unlike those for the [557] and [561] states of YbF. It can be concluded that the state at 17908 $cm^{-1}$ is the relatively unperturbed $\tilde{A}\,^2\Pi_{1/2}(1,0,0)$ level. The short radiative lifetimes ($\approx$ 22 ns), near-diagonal fluorescence and the rotational analyses[14] for the $\tilde{A}\,^2\Pi_{1/2}(0,0,0)$ and $\tilde{A}\,^2\Pi_{1/2}(1,0,0)$ states indicates a predominant $4f^{14}6p^1$ ($Yb^+$) character for both levels. The observed long progression in Yb-OH stretching, $\nu_1''$, in the DLIF spectrum and the relatively long radiative lifetime (124 ± 2ns) of the [17.73]0.5 state is evidence that the dominant configuration for this state is $\left([\mathrm{Xe}]4f^{13}\right)_{\mathrm{Yb}^+}\sigma^2_{\mathrm{Yb}^+(6s6p)}$. No other states of this nature have been observed at lower energy suggesting that the excited state of the 17730 $cm^{-1}$ band is [17.73]0.5(0,0,0).

Assignment of the [17.33], [17.637], [17.643], [17.68], [17.71] and [17.90] states is more difficult. The DLIF spectra resulting from the excitation of these six excited electronic states exhibit numerous transitions that are nominally vibrationally forbidden. Such bands are parallel polarized although the $\tilde{A}\,^2\Pi_{1/2}-\tilde{X}\,^2\Sigma^+$ electronic transition moment is perpendicular. These transitions appear as a result of coupling through the bending vibrations between electronic states whose lambda values differ by one unit (i.e. $\tilde{A}\,^2\Pi_{1/2}$ state coupling with the $^2\Sigma$ and $^2\Delta$ electronic states)[40]. Levels from $\tilde{A}\,^2\Pi_{1/2}(0,1^1,0)$, and possibly some components of the $\tilde{A}\,^2\Pi_{1/2}(0,2^l,0)$ vibronic states, will have energies between those of the $\tilde{A}\,^2\Pi_{1/2}(0,0,0)$ and $\tilde{A}\,^2\Pi_{1/2}(1,0,0)$ levels. The $\tilde{A}\,^2\Pi(0,1^1,0)$ and $\tilde{A}\,^2\Pi(0,2^l,0)$ vibronic states are subject to both Renner-Teller and spin-orbit interactions[41] and are characterized by the approximately good quantum numbers after accounting for these interactions. In the absence of the electronic spin, the projection of the total angular momentum on to the symmetry axis is $\hbar K = \hbar\left(\Lambda + l\right)$, where $\Lambda$ and $l$ are the quantum numbers associated with the projection of the total electronic orbital and vibrational angular momenta on the symmetry axis. The large spin-orbit coupling expected for the $\tilde{A}^2\Pi$ state of YbOH suggests that the total electron spin angular momentum, $\vec{S}$, is quantized in the molecular frame. The projection of total electron spin angular momentum on the symmetry axis is $\Sigma$. Consequently, the vibronic energy levels will be approximately those of a Hund's case (a)-type angular momentum coupling scheme where the quantum number $P = \Lambda + \Sigma + l$ associated with the projection of the total vibronic angular momentum on the symmetry axis is conserved. Standard practice[37] is to label the Renner-Teller and spin-orbit vibronic states by ${}^{2S+1}|K|_{|P|}$. Therefore, the expected Hund's case (a)-type levels from $\tilde{A}^2\Pi_{1/2}(0,1^1,0)$ and $\tilde{A}^2\Pi_{1/2}(0,2^l,0)$ vibronic states are:

$$\tilde{A}^2\Pi(0,1^1,0) \rightarrow \mu^2\Sigma,\ \kappa^2\Sigma,\ {}^2\Delta_{3/2} \text{ and } {}^2\Delta_{5/2} \tag{5}$$

and

$$\tilde{A}^2\Pi(0,2^l,0) \rightarrow {}^2\Phi_{5/2},\ {}^2\Phi_{7/2},\ \mu^2\Pi_r \text{ and } \kappa^2\Pi_i\,. \tag{6}$$

The lower and upper energy vibronic states with the same symmetry are distinguished by 'μ' and 'κ', respectively. The large spin-orbit interaction ($A\approx1350$ cm$^{-1}$) places the $\kappa^2\Sigma$ and ${}^2\Delta_{3/2}$ levels of the $\tilde{A}^2\Pi_{1/2}(0,1^1,0)$ state and the ${}^2\Phi_{7/2}$ and $\kappa^2\Pi_i$ levels of the $\tilde{A}^2\Pi(0,2^l,0)$ state at energies outside the probed spectral range[37,41,42]. Assuming that the excited state bending frequency, $\omega_2'$, is similar to that of the $\tilde{X}^2\Sigma^+$ state of 329 cm$^{-1}$, then the $\mu^2\Sigma$ level of the $\tilde{A}^2\Pi_{1/2}(0,1^1,0)$ state should be near 17650 cm$^{-1}$ (=$T_{00}+\omega_2'$) and the ${}^2\Delta_{5/2}$ level only slightly higher. The [17.643] state, which fluoresces with high efficiency to the to the $\tilde{X}^2\Sigma^+(0,1^1,0)$ level (Figure 7), is likely to have a dominant $\tilde{A}^2\Pi_{1/2}(0,1^1,0)\mu^2\Sigma$ character. This is supported by the observed short lifetime (35± 6 ns) and the predicted branching ratios (see below). The $\tilde{A}^2\Pi_{1/2}(0,1^1,0)\mu^2\Sigma - \tilde{X}^2\Sigma^+(0,0,0)$ transition is expected[40] to be more intense than the $\tilde{A}^2\Pi_{1/2}(0,1^1,0)^2\Delta_{5/2} - \tilde{X}^2\Sigma^+(0,0,0)$ transition, though both are vibronically induced. The [17.68] and [17.71] excited states associated with the relatively intense $[17.68]-\tilde{X}^2\Sigma^+(0,0,0)$ band and the much weaker $[17.71]-\tilde{X}^2\Sigma^+(0,0,0)$ band (Figure 6) are most likely highly mixed with the near [17.73]0.5, $\left([\mathrm{Xe}]4f^{13}\right)_{\mathrm{Yb}^+}\sigma^2_{\mathrm{Yb}^+(6s6p)}$ state. The [17.68] state was assigned as nominally the $\tilde{A}^2\Pi_{1/2}(0,1^1,0)\mu^2\Sigma$ level in the study of the high temperature sample[12].

The [17.33] and [17.90] states, neither of which were reported in the previous study[14], are particularly interesting. The excitation spectra for the $[17.33]-\tilde{X}^2\Sigma^+(0,0,0)$ and $[17.90]-\tilde{X}^2\Sigma^+(0,0,0)$ band are approximately a factor of four weak than those for the adjacent $\tilde{A}^2\Pi_{1/2}(0,0,0)-\tilde{X}^2\Sigma^+(0,0,0)$ and $\tilde{A}^2\Pi_{1/2}(1,0,0)-\tilde{X}^2\Sigma^+(1,0,0)$ bands. The [17.33] state, which is approximately 10 $cm^{-1}$ higher than the $\tilde{A}^2\Pi_{1/2}(0,0,0)$ state, and the [17.90] state, which is approximately 10 $cm^{-1}$ lower than the $\tilde{A}^2\Pi_{1/2}(1,0,0)$ state, have very nearly identical DLIF spectra and fluorescent lifetimes as the nearby and $\tilde{A}^2\Pi_{1/2}(0,0,0)$ and $\tilde{A}^2\Pi_{1/2}(1,0,0)$ levels. The very diagonal nature of the DLIF spectra for these states implies that potential energy surfaces for the excited states are very nearly identical to those of the $\tilde{X}^2\Sigma^+$ state. The short radiative lifetimes (≈22 ns) for all four excited states implies that that they are all associated with $Yb^+$-centered atomic like transitions. The exact nature of the [17.33] and [17.90] states is a mystery at this point. One very speculative assignment is that YbOH in the $\tilde{A}^2\Pi_{1/2}$ is slightly bent and that two bands are associated with excitation to the $K_a$=0 and $|K_a|$=1 levels of an asymmetric rotor. Under this assumption the 17323 and 17332 cm$^{-1}$ bands would be assigned as the $\tilde{X}\,^2\Sigma^+(0,0,0) \rightarrow \tilde{A}\,^2A'(K_a = 0)$ and $\tilde{X}\,^2\Sigma^+(0,0,0) \rightarrow \tilde{A}\,^2A'(|K_a| = 1)$ transitions, respectively. The observed relative intensities, with the higher energy band at 17332 cm$^{-1}$ band being much weaker, would be consistent with this assignment. A similar assignment for the weak 17900 and more intense 17908 cm$^{-1}$ bands would require the highly unusual case of the $|K_a|$=1 levels being lower in energy than the $K_a$=0 levels. The switching of $K_a$ ordering has been observed in the bending vibrational levels of quasilinear molecules[43] but not, to our knowledge, in stretching modes.

### C. Prediction of branching ratios

The measured $b_{iv',fv''}$ values for $\tilde{A}\,^2\Pi_{1/2}(0,0,0) \rightarrow \tilde{X}\,^2\Sigma^+$ (0,0,0), (1,0,0), (0,2$^0$,0) and (2,0,0) transitions are 89.73%, 9.74%, 0.27% and 0.26%, all with estimated errors of 0.05% , which compared favorably with the previously predicted[1] values of 86.73%, 11.73% ,0.10% and 0.13%. In the present study, the branching ratios for the $\tilde{A}\,^2\Pi_{1/2}(0,0,0)$ state, as well as those for the $\tilde{A}\,^2\Pi_{1/2}(1,0,0)$ and $\tilde{A}\,^2\Pi_{1/2}(0,1^1,0)$ states, were also predicted assuming that these levels are unperturbed and that stretching and bending potentials for the $\tilde{A}\,^2\Pi_{1/2}$ and $\tilde{X}\,^2\Sigma^+$ states are harmonic. Assuming the Born-Oppenheimer approximation the branching ratios are given by,

$$b_{iv',fv''} = \frac{\nu^3 q_{v'v''}}{\sum_i \nu_i^3 q_{v'v''}}, \tag{7}$$

where $q_{v'v''}$ are the Franck-Condon factors (FCFs) and ν is the emission frequency. This assumes that vibronic coupling interactions (i.e. Coriolis, Renner-Teller, Fermi, etc.) and spin-orbit interactions are negligible and the total wave function can be written as the product of an electronic and vibrational wave function. Under these assumptions the relative intensities are proportional to the product of the two-dimensional FCF of the $\sigma$-symmetry stretching modes ($\nu_1$ and $\nu_3$) and a FCF for the one-dimensional $\pi$-symmetry bending mode ($\nu_2$):

$$\mathrm{FCF} = \left|\left\langle \nu_1', \nu_3' \middle| \nu_1'', \nu_3'' \right\rangle\right|^2 \left|\left\langle \nu_2' \middle| \nu_2'' \right\rangle\right|^2 . \tag{8}$$

In the present study the one-dimensional $\pi$-symmetry bending mode FCFs are evaluated using the analytical formula for non-displaced harmonic oscillators[44]. Evaluation of the two-dimensional FCF is more problematic. The previous prediction[1] of FCFs and $b_{iv',fv''}$ values followed the procedure of Sharp and Rosenstock[45] to evaluate the two dimensional integrals of Eq. 8. Here alternative closed-form formulas[46,47] for the two-dimensional FCFs are employed similar to what was carried out for modelling the SrOH two-dimensional FCFs[31]. In the SrOH study only emission from the lowest vibrational level of the excited electronic state (i.e. $\tilde{A}^2\Pi_{1/2}(0,0,0)$) was modeled and the formula for the two-dimensional FCF, $\left|\left\langle \tilde{A}^2\Pi, \nu_1 = 0, \nu_3 = 0 \middle| \tilde{X}^2\Sigma^+, \nu_1, \nu_3 \right\rangle\right|^2$, derived by Chang[46] was employed. Modelling the relative intensities of the present YbOH DLIF spectra requires using formula for FCFs of the more general form $\left|\left\langle \tilde{A}^2\Pi, \nu_1 \neq 0, \nu_3 \neq 0 \middle| \tilde{X}^2\Sigma^+, \nu_1, \nu_3 \right\rangle\right|^2$, which have recently been derived by Sattasathuchana et al.[47]. The analytical solutions such as those derived by Chang[46] and Sattasathuchana et al.[47] for the two-dimensional FCFs have some advantages over those of Sharp and Rosenstock[45] because they are relatively easy to code, are exact and free from convergence problems. All three approaches assume a harmonic motion and account for the change in normal modes upon excitation (i.e. Duschinsky effect). The Sharp and Rosenstock[45] approach treats the Duschinsky effect by assuming the same internal coordinates for the ground and excited states whereas the methods by Chang[46] and Sattasathuchana et al.[47] employ the more accurate method of using the Cartesian displacement coordinates common to both electronic states[48]. This approach requires relating the normal coordinates of the $\tilde{X}\,^2\Sigma^+$ state, **Q**( $\tilde{X}\,^2\Sigma^+$), to those of the $\tilde{A}\,^2\Pi$ states, **Q**( $\tilde{A}\,^2\Pi$ ) :

$$\mathbf{Q}(\tilde{X}\,^2\Sigma^+) = \mathbf{J}\mathbf{Q}(\tilde{A}\,^2\Pi) + \mathbf{D} \; . \qquad (9)$$

In Eq. 9, **D** is the vector of geometry displacements given in terms of the normal coordinates of the ground state and $\mathbf{J}$ is the Duschinsky rotation matrix. For the linear-to-linear transition studied here the $\mathbf{J}$ rotation matrix associated with the two σ-type stretching modes is a unit matrix. Details of the prediction are found in Appendix A.

The determined two dimensional $\left|\left\langle \tilde{A}\,^2\Pi, v_1, v_3 \middle| \tilde{X}\,^2\Sigma^+, v_1, v_3 \right\rangle\right|^2$ and one dimensional $\left|\left\langle \tilde{A}\,^2\Pi, v_2 \middle| \tilde{X}\,^2\Sigma^+, v_2 \right\rangle\right|^2$ FCFs are presented in Table 3. Also presented are the predicted and observed $b_{iv',fv''}$ values. The agreement for the DLIF spectra for emission from the $\tilde{A}\,^2\Pi_{1/2}(0,0,0)$ and $\tilde{A}\,^2\Pi_{1/2}(1,0,0)$ states are reasonable. The predicted branching ratios for the $\tilde{A}\,^2\Pi_{1/2}(0,1^1,0)$ levels most closely match those for the [17.643] level.

### D. Relevance to laser cooling

The measurements performed here can be used to determine the optimal scheme to achieve rapid photon cycling for laser cooling of YbOH. Typical molecular laser cooling experiments involve ~$10^4$ - $10^5$ photon scatters per molecule in order to cool and trap molecules at < mK temperature. Achieving this level of photon cycling requires “repumping” all population that decays to levels with probability >$10^{-4}$. Practical considerations motivate directing these repumping lasers through several different excited vibronic levels in order to maximize the scattering rate, and therefore increase the capture velocity of the cooling laser beams. These two

concerns—closing off vibrational loss channels and maximizing the photon scattering rate— are often in competition because higher-lying excited vibronic states tend to have less diagonal Franck-Condon factors.

We use a Markov chain model to predict the average number of photons scattered before molecules are optically pumped into a dark vibrational state. We make the reasonable assumption that the decays from a given excited level to $\tilde{X}\,^2\Sigma^+(4,0,0)$ are about half the size of the corresponding decay to $\tilde{X}\,^2\Sigma^+(3,0,0)$, which is the measured ratio of the DLIF spectrum resulting from the [17.73]0.5 level. Several viable repumping schemes are displayed in Figure 14. Figure 14(a) shows a simple scheme in which all repumping transitions are driven through the $\tilde{A}\,^2\Pi_{1/2}(0,0,0)$ state. Based on our measurements, this scheme will allow ~4,000 photon scatters before optical pumping into a dark state. While this maximizes the number of photons that can be scattered for a given number of lasers, it decreases the attainable scattering rate by a factor of ~16 (Ref. [49]). Our measurements also indicate that the [17.73]0.5 and $\tilde{A}\,^2\Pi_{1/2}(1,0,0)$ states will be useful for "two-step" repumping methods. Figures 14(b) and 14(c) show how these states can be used to provide auxiliary repumping routes. The Markov chain model predicts that both schemes will allow ~3,000 photon scatters before optical pumping into the dark vibrational levels. However, either scheme increases the scattering rate by a factor of ~2.3 relative to scheme (a). Because the capture velocity will scale with this scattering rate, such a tradeoff is favorable.

Several of the excited electronic states are not directly useful for laser cooling due to their non-diagonal Franck-Condon factors. However, these states are useful for the spectroscopic task of locating losses from the optical cycle. This is because they allow optical pumping into relatively high-lying vibrational levels in the $\tilde{X}\,^2\Sigma^+$ manifold, e.g. $\tilde{X}\,^2\Sigma^+(4,0,0)$ or $\tilde{X}\,^2\Sigma^+(0,4^0,0)$. A pump-probe style experiment can then be used to determine repumping pathways with rotational resolution. Furthermore, the finding that the [17.68] and [17.643] levels couple strongly to both the $\tilde{X}\,^2\Sigma^+(0,0,0)$ and $\tilde{X}\,^2\Sigma^+(0,1^1,0)$ levels is important in that it will allow efficient optical pumping into the metastable bending mode proposed for the ultimate electron EDM measurement in trapped YbOH molecules.

## VII. SUMMARY

The measurements reported here will assist in implementation of efficient photon cycling and repumping schemes required for laser cooling and trapping and were already exploited in the recent demonstration of one dimensional laser-cooling of YbOH to temperatures around 10 uK[19] . Although understanding the excited state distribution is problematic, the vibronic energy pattern for $\tilde{X}\,^2\Sigma^+$ appears to be free of local perturbations up to an energy of 2100 cm$^{-1}$. The $\tilde{X}\,^2\Sigma^+(0,1^1,0)$ state, which is the target for the EDM measurements[1], can be state selectively populated by exciting the relatively intense $\tilde{X}\,^2\Sigma^+(0,0,0)\rightarrow[17.68]$ transition near 17681 cm$^{-1}$ and/or the weaker $\tilde{X}\,^2\Sigma^+(0,0,0)\rightarrow[17.643]$ transition. The strongly vibronically mixed [17.68] excited state, which exhibits some $\tilde{A}\,^2\Pi_{1/2}(0,1^1,0)$ character, fluoresces with relatively high probability ($b_{iv',fv''}$=21.8 %) to the desired $\tilde{X}\,^2\Sigma^+(0,1^1,0)$ state. Similarly, the [17.643] excited state preferentially fluoresces ($b_{iv',fv''}$=73.1 %) to the desired $\tilde{X}\,^2\Sigma^+(0,1^1,0)$ state.

The *ab initio* vibrational levels for the electronic ground state of YbOH obtained from DVR calculations using relativistic coupled-cluster potential energy surface agree quite well with measurement. The predicted location of unobserved levels will assist in future spectroscopic studies. It is of significant interest to generalize the present computational scheme to include treatment of Renner-Teller effects and to enable accurate *ab initio* calculations for vibronic levels of $\tilde{A}^2\Pi$ states.

**Acknowledgements**

The research at Arizona State University was supported by a grant from the Heising-Simons Foundation (Grant 2018-0681). The authors thank Prof. Michael Morse (Chemistry Department. University of Utah) for the use of a cw-dye laser system and Prof. Nicholas Hutzler (Division of Physics, Mathematics, and Astronomy, California Institute of Technology) for his insightful comments.

**Supporting Information Available:**

Details of the DVR prediction are provided. The harmonic vibration frequencies and dimensionless reduced normal coordinates (in Bohr) as well as equilibrium structure in Cartesian coordinate (in Bohr) of YbOH used in the DVR calculation are shown in Table S1 and the equilibrium structure in Table S2. The coefficients of six-order polynomial analytical potential energy function obtained by fitting the *ab initio* energies are presented in Table S3. This material is available free of charge via the Internet at http://pubs.acs.org

## Appendix A: The two-dimensional stretching Franck-Condon factors.

The displacement vector of the Duschinsky transformation is given by:

$$\mathbf{D}=(\mathbf{L}^{-1}(\tilde{A}^2\Pi_r)\cdot\mathbf{B}(\tilde{A}^2\Pi_r))\cdot(\mathbf{R}_{eq}(\tilde{A}^2\Pi_r)-\mathbf{R}_{eq}(\tilde{X}^2\Sigma^+)) \quad , \qquad \text{(A1)}$$

where $\mathbf{R_{eq}}$ is the vectors of equilibrium Cartesian coordinates in a center of mass where the atoms lie along the z-axis. The **B** and **L** matrices for the ground and excited states are required to account for the Duschinsky effect[45–48] (i.e. mode mixing). The initial step in the FCF prediction is a normal coordinate analyses using the **GF** matrix approach, which is well documented by Wilson Decius & Cross[50]. The **GF** matrix analysis provides the requisite **B** and **L** matrices that relate the normal coordinates, **Q,** to the internal symmetry coordinate, **S**, and then those to the Cartesian displacement coordinates **X.** The expressions for the elements of the **B, G** and **F** matrices can be found in the SrOH manuscript[51]. The expression for the **F** matrix is:

$$\mathbf{F}=\left|\begin{array}{ccccc} & Yb-O & O-H & (Yb-O-H)^{xz} & (Yb-O-H)^{yz} \\ Yb-O & f_{11} & f_{12} & 0 & 0 \\ O-H & f_{12} & f_{22} & 0 & 0 \\ (Yb-O-H)^{xz} & 0 & 0 & f_{33} & 0 \\ (Yb-O-H)^{yz} & 0 & 0 & 0 & f_{33} \end{array}\right| \qquad \text{(A2)}$$

The parameters used for generating the **G** and **F** matrices for the $\tilde{X}^2\Sigma^+$ and $\tilde{A}^2\Pi$ states of YbOH are given in Table A1. The O-H bond distances, $r_{O\text{-}H}$, for the $\tilde{X}^2\Sigma^+$ and $\tilde{A}^2\Pi$ states were taken as that measured for the $\tilde{X}^2\Sigma^+$ state of BaOH[52]. The O-H stretch stretching force constants, $f_{33}$, for the $\tilde{X}^2\Sigma^+$ and $\tilde{A}^2\Pi$ states were taken as 7.836 mdyne/Å, which was obtained using the *ab initio* predicted[53] ground $X^1\Sigma^+$ harmonic frequency for $OH^-$ (3746.6 $cm^{-1}$). The stretch-stretch coupling force constant, $f_{13}$, was constrained to the *ab initio* predicted[54] value (0.0677 mdyne/Å) for the ground $\tilde{X}\,^2\Sigma^+$ state of CaOH. The Yb-O stretching force constants, $f_{11}$, for the $\tilde{X}\,^2\Sigma^+$ and $\tilde{A}\,^2\Pi$ states were taken as 2.60 mdyne/Å and 3.00 mdyne/Å , respectively. These values were obtained by fixing $f_{33}$ and $f_{13}$ and varying $f_{11}$ to reproduce the observed $\omega_3$(Yb-O stretch) values of 529 $cm^{-1}$ and 585 $cm^{-1}$ for the $\tilde{X}\,^2\Sigma^+$ and $\tilde{A}\,^2\Pi$ states. Similarly, the Yb-O-H bending force constants, $f_{22}$, for the $\tilde{X}\,^2\Sigma^+$ and $\tilde{A}\,^2\Pi$ states were obtained by reproducing observed $\omega_2$(Yb-O-H bend) values of 328 and 357 $cm^{-1}$.

**Table A1. Force constant and bond lengths used in the normal coordinate analysis.**

| Property | $\tilde{X}\,^2\Sigma^+$ | $\tilde{A}\,^2\Pi$ |
|---|---|---|
| Yb-O stretch, $f_{11}$ [a] | 2.60 | 3.13 |
| Yb-O-H bend, $f_{33}$ | 0.0500 | 0.058 |
| O-H stretch , $f_{22}$ [b] | 7.939 | 7.939 |
| stretch-stretch $f_{12}$ | 0.0677 | 0.0677 |
| $r_{\text{Yb-O}}$(Å)[c] | 2.0397 | 2.006 |
| $r_{\text{O-H}}$ (Å)[c] | 0.927 | 0.927 |
| $\omega_3$(O-H stretch)[b] | 3746.6 | 3746.6 |
| $\omega_2$(Yb-O-H bend) | 328 | 357 |
| $\omega_1$(Yb-O stretch) | 529 | 585 |

a) Units: $f_{11}$, $f_{33}$, and $f_{13}$ in mdyne/Å; $f_{22}$ mdyne/(Å radian).
b) From *ab initio* predicted[53] ground $X^1\Sigma^+$ harmonic frequency for $OH^-$ (3746.6 $cm^{-1}$)
c) Constrained to the predicted value for CaOH[54].

The $\mathbf{L}$ matrices are related to the eigenvectors of the **GF** matrices, $\mathbf{V}$, by a normalization matrix $\mathbf{N}$:

$$\mathbf{L}=\mathbf{V}\cdot\mathbf{N}=\mathbf{V}\cdot\left[\mathbf{V}^{-1}\mathbf{G}(\mathbf{V}^{\mathrm{T}})^{-1}\right]^{1/2}, \qquad \text{(A3)}$$

where $\mathbf{N}$ is chosen such that $\mathbf{L}\cdot\mathbf{L}^T=\mathbf{G}$. Using information of Table 1A, the calculated **V** and **L** matrices are for σ-type stretching modes are:

$$\mathbf{V}(\tilde{X}\,^2\Sigma^+)\approx\begin{vmatrix} & Q_1 & Q_2 \\ \Delta r_{O-H} & 0.9982 & 0.0111 \\ \Delta r_{Yb-O} & -0.0598 & 0.9999\end{vmatrix} \qquad \mathbf{V}(\tilde{A}\,^2\Pi)\approx\begin{vmatrix} & Q_1 & Q_2 \\ \Delta r_{O-H} & 0.9982 & 0.0152 \\ \Delta r_{Yb-O} & 0.0601 & 0.9999\end{vmatrix} \qquad \text{(A4)}$$

$$\mathbf{L}(\tilde{X}\,^2\Sigma^+) \approx \begin{vmatrix} & Q_1 & Q_2 \\ \Delta r_{O-H} & 1.0270 & 0.0028 \\ \Delta r_{Yb-O} & -0.0616 & 0.2539 \end{vmatrix} \qquad \mathbf{L}(\tilde{A}\,^2\Pi) \approx \begin{vmatrix} & Q_1 & Q_2 \\ \Delta r_{O-H} & 1.0270 & 0.0039 \\ \Delta r_{Yb-O} & -0.0618 & 0.2539 \end{vmatrix} \quad . \quad \text{(A5)}$$

As expected $\mathbf{V}$ and $\mathbf{L}$ are nearly identical for the $\tilde{X}\,^2\Sigma^+$ and $\tilde{A}\,^2\Pi$ states. Substitution of $\mathbf{L}$ and $\mathbf{B}$ into Eq. 7 produces a unit $\mathbf{J}$ rotation matrix, as expected because the normal coordinates $Q_1$ and $Q_3$ are parallel for the three states. Substitution of $\mathbf{L}, \mathbf{B}$ and the bond lengths of Table A1 into Eq. 8 gives:

$$\mathbf{D}(\tilde{X}\,^2\Sigma^+, \tilde{A}\,^2\Pi) \approx \begin{vmatrix} -0.00069 \\ 0.13179 \end{vmatrix} \quad . \qquad \text{(A6)}$$

In our calculations we used a generalized version of the expression derived by Chang[46] for the special case of the Franck Condon factors of the form $\left|\left\langle 0,0 \middle| \nu_1, \nu_2 \right\rangle\right|^2$. In particular we evaluated the integrals $\left\langle \tilde{A}\,^2\Pi, \nu_1 \neq 0, \nu_3 \neq 0 \middle| \tilde{X}\,^2\Sigma^+, \nu_1, \nu_3 \right\rangle \left\langle \tilde{A}\,^2\Pi, \nu_2 \middle| \tilde{X}\,^2\Sigma^+, \nu_2 \right\rangle$, where the stretching and bending modes have been factored due to their different symmetry. Following on the derivation by Chang[46], the generalized expression for the two-dimensional FCF in the harmonic approximation is:

$$|\langle v_1' v_2' | v_1 v_2 \rangle|^2 = (N_{v_1 v_2; v_1' v_2'} E H_{v_1 v_2; v_1' v_2'})^2 \qquad \text{(A7)}$$

where $$N_{v_1 v_2; v_1' v_2'} = \frac{\sqrt{\alpha_1 \alpha_2 \alpha_1' \alpha_2'}}{2^{v_1+v_2+v_1'+v_2'} v_1! v_2! v_1'! v_2'! A_1 A_2} \quad , \qquad \text{(A8)}$$

$$E = \exp(-\tfrac{1}{2}\alpha_1' D_1^2 - \tfrac{1}{2}\alpha_2' D_2^2 + A_1 C_1^2 + A_2 C_2^2) \quad , \qquad \text{(A9)}$$

$$\alpha_i = \frac{\omega_i}{\hbar}, \qquad \text{(A10)}$$

and

$$H_{v_1 v_2; v_1' v_2'} = \sum_{k_1=0}^{v_1} \sum_{k_2=0}^{v_1-k_1} \sum_{k_3=0}^{v_2} \sum_{k_4=0}^{v_1'} \sum_{k_5=0}^{v_1'-k_4} \sum_{k_6=0}^{v_2'} \sum_{k_7=0}^{v_2'-k_6} \binom{v_1}{k_1} \binom{v_1-k_1}{k_2} \binom{v_2}{k_3} \times \binom{v_1'}{k_4} \binom{v_1'-k_4}{k_5} \binom{v_2'}{k_6} \binom{v_2'-k_6}{k_7} H_{v_1-k_1-k_2}(b_1)\, H_{v_2-k_3}(b_2)\, H_{v_1'-k_4-k_5}(b_3)\, H_{v_2'-k_6-k_7}(b_4) \times F_1^{k_1} F_2^{k_2} F_3^{k_3} F_4^{k_4} F_5^{k_5} F_6^{k_6} F_7^{k_7} (k_1 + k_4 + k_6 - 1)!!\, (k_2 + k_3 + k_5 + k_7 - 1)!! \quad , \qquad \text{(A11)}$$

with the restriction that $k_1 + k_4 + k_6$ even and $k_2 + k_3 + k_5 + k_7$ even. We define:

$F_4 = \left(\frac{2}{A_1}\right)^{\frac{1}{2}} a_4$, (A12); $F_5 = \left(\frac{2}{A_2}\right)^{\frac{1}{2}} a_5$ (A13); $F_6 = \left(\frac{2}{A_1}\right)^{\frac{1}{2}} a_6$ (A14); $F_7 = \left(\frac{2}{A_2}\right)^{\frac{1}{2}} a_7$ (A15);

$b_3 = \sqrt{\alpha_1'}(B_1 C_2 J_{11} - C_1 J_{11} - C_2 J_{12} + D_1)$ (A16); $b_4 = \sqrt{\alpha_2'}(B_1 C_2 J_{21} - C_1 J_{21} - C_2 J_{22} + D_2)$ (A17);

$a_4 = \sqrt{\alpha_1'} J_{11}$ (A18); $a_5 = \sqrt{\alpha_1'}(-B_1 J_{11} + J_{12})$ (A18); $a_6 = \sqrt{\alpha_2'} J_{21}$ (A19);

$a_7 = \sqrt{\alpha_2'}(-B_1 J_{21} + J_{22})$ (A20);

The definitions of $b_1, b_2, F_1, F_2, F_3, A_1, A_2, B_1, C_1$, and $C_2$ can be found in the original derivation[46] for the $\left|\langle 0,0|\nu_1,\nu_2\rangle\right|^2$ factors. The two-dimensional FCFs, $\left|\langle \tilde{A}\,^2\Pi,\nu_1,\nu_3 | \tilde{X}\,^2\Sigma^+,\nu_1,\nu_3\rangle\right|^2$, presented in Table 3 were obtained using $D_1$ and $D_2$ of Eq. A6 , the harmonic frequencies $\omega_1(\tilde{X}\,^2\Sigma^+)$, $\omega_2(\tilde{X}\,^2\Sigma^+)$, $\omega_1(\tilde{A}\,^2\Pi_{1/2})$ and $\omega_2(\tilde{A}\,^2\Pi_{1/2})$ of Table A1 and Eq. A7.

**Figure Captions:**

**Fig. 1.** The energies levels and associated state assignments for the dominant spectral features in the laser excitations spectrum of a supersonically cooled $^{174}$YbOH sample in the 17300 cm$^{-1}$ to 17950 cm$^{-1}$ spectral range. The vibrational quantum numbers associated with the $\tilde{A}^2\Pi_{1/2}$ and $\tilde{X}^2\Sigma^+$ are given in parentheses. The numbers next to the quantum numbers for the $\tilde{X}^2\Sigma^+$ state are the energies in wavenumbers. The $\tilde{A}^2\Pi_{1/2}(0,0,0)-\tilde{X}^2\Sigma^+(0,0,0)$, $\tilde{A}^2\Pi_{1/2}(1,0,0)-\tilde{X}^2\Sigma^+(0,0,0)$, $\tilde{A}^2\Pi_{1/2}(0,0,0)-\tilde{X}^2\Sigma^+(1,0,0)$ and [17.73]0.5-$\tilde{X}^2\Sigma^+(0,0,0)$ transition have been recorded at high resolution and analyzed[14,15].

**Fig. 2.** The 2D spectrum in the 17315 to 17385 cm$^{-1}$ spectral range. The horizontal axis is the excitation wavelength of the pulsed dye lase and the vertical axis is the shift in wavelength (nm) of the dispersed the fluorescence relative to the excitation wavelength.

**Fig. 3.** Excitation spectra extracted by vertical integration of the signal along the horizontal slices of the 2D spectrum of Figure 1 and the predicted LIF spectrum in the17315 to 17385 cm$^{-1}$ range. A) on- resonance ("Ex1"); B) Stokes shifted by one quantum of Yb-OH stretch $\nu_1''$ ("Ex2"); C) anti-Stokes shifted by one quantum bending $\nu_2''$ ("Ex3"); D) the predicted LIF excitation spectrum based upon the previous analysis[14].

**Fig. 4.** Left: The dispersed laser induced fluorescence spectra resulting from pulsed dye laser excitation of the band heads at 17323, 17332, 17345, and 17375 cm$^{-1}$. The features marked as "Ex" is the emission occurring at the laser excitation wavelength. The numbers above the spectral features are the measured shifts in wavenumber (cm$^{-1}$) relative to the laser. Right: The energy levels and associated quantum number assignments.

**Fig. 5.** The 2D spectrum in the 17625 to 17750 cm$^{-1}$ spectral range. The band near 17300 cm$^{-1}$ is the [17.73]0.5-$\tilde{X}\,^2\Sigma^+(0,0,0)$ transition and has been recorded at high resolution and analyzed[13]. The weak, non-horizontal, emission shifted by approximately -11.5 nm (≈365 cm$^{-1}$) at 17630 nm is an artifact of amplified stimulated emission (ASE) of the pulsed dye laser. The ASE is exciting the very strong $6s^2\ ^1S_0 \rightarrow 6s6p\ ^3P_1$ transition of Yb(I) at 17992.007 cm$^{-1}$ giving rise to an emission that is shift from the laser wavelength.

**Fig. 6.** Excitation spectra extracted by vertical integration of the signal along the horizontal slices of the 2D spectrum of Figure 5 in the 17625 to 17750 cm$^{-1}$ range. A) on-resonance ("Ex1"); B) Stokes shifted by one quantum of bending, $\nu_2''$ ("Ex2"); C) anti-Stokes shifted by one quantum of Yb-OH stretch, $\nu_1''$ ("Ex3").

**Fig. 7.** Left: The dispersed laser induced fluorescence spectra resulting from pulsed dye laser excitation of the band heads at 17637, 17643, 17681, 17708 and 17730 cm$^{-1}$. The feature marked as "Ex" is the emission occurring at the laser excitation wavelength. The negative spike in DLIF2 is due to pulsed dye laser amplified stimulated emission (ASE) and an imperfect background subtraction. The numbers above the spectral features are the measured shifts in wavenumber (cm$^{-1}$) relative to the laser. Right: The energy levels and associated quantum number assignments.

**Fig. 8.** Bottom: The 2D spectrum in the 17880 to 17920 cm$^{-1}$ spectral range. Top: Excitation spectrum extracted by vertical integration of the signal along the horizontal slice of the 2D spectrum which is Stokes shifted by one quantum of bending, $\nu_2''$ . The band near 17908 cm$^{-1}$ is the $\tilde{A}\,^2\Pi_{1/2}(1,0,0) - \tilde{X}\,^2\Sigma^+(0,0,0)$ transition and has been recorded at high resolution and analyzed[14].

**Fig. 9.** Left: The dispersed laser induced fluorescence spectra resulting from pulsed dye laser excitation of the band heads at 17900 and 17908 cm$^{-1}$. The features marked as "Ex" are emission

occurring at the laser excitation wavelength. The "*" marked feature is due to emission from the excited $6s^2\ ^1S_0 \rightarrow 6s6p\ ^3P_1$ transition of Yb(I) at 17992.007 $cm^{-1}$ excited by pulsed dye laser amplified stimulated emission(ASE) and an imperfect background subtraction. The numbers above the spectral features are the measured shifts in wavenumber ($cm^{-1}$) relative to the laser. Right: The energy levels and associated quantum number assignments.

**Fig. 10.** Fluorescence decay data for the $\tilde{A}^2\Pi_{1/2}(0,0,0)$, [17.33], $\tilde{A}^2\Pi_{1/2}(1,0,0)$ and [17.90] states of YbOH. The dashed lines are predicted decays using the optimized lifetimes. The $\tilde{A}^2\Pi_{1/2}(0,0,0)$, [17.33], $\tilde{A}^2\Pi_{1/2}(1,0,0)$ and [17.90] states exhibit very diagonal ($\Delta$v=0) fluorescence.

**Fig. 11.** Fluorescence decay data for the [17.637], [17.642], [17.68], and [17.73]0.5 states of YbOH. The dashed lines are predicted decays using the optimized lifetimes. The [17.637], [17.642], [17.68]and [17.73]0.5 states exhibit very non-diagonal ($\Delta$v$\neq$0) fluorescence.

**Fig. 12.** Fluorescence decay data resulting from exciting the band at 17345 $cm^{-1}$, which is nominally the $\tilde{A}^2\Pi_{1/2}(0,1^1,0)-\tilde{X}^2\Sigma^+(0,1^1,0)$ transition. The band at 17345 $cm^{-1}$ is overlapped with high-$J$ transition of the $\tilde{A}^2\Pi_{1/2}(0,0,0)-\tilde{X}^2\Sigma^+(0,0,0)$ and $[17.33]-\tilde{X}^2\Sigma^+(0,0,0)$ bands. Decay curves obtained by monitoring the on-resonance emission exhibits a short lifetime (21±4 ns) similar to that of the resulting from exciting the $\tilde{A}^2\Pi_{1/2}(0,0,0)$ and [17.33] states. The anti-Stokes emission shifted by one quantum of Yb-OH bending, $\nu_2''$, exhibits a longer lifetime (110±15 ns) and is due to emission from the [17.68] level which is nominally $\tilde{A}^2\Pi_{1/2}(0,1^1,0)$.

**Fig. 13.** A comparison of the dispersed laser induced fluorescence spectra resulting from pulsed dye laser excitations of the band heads at 17323, 17730 and 17908 $cm^{-1}$ and continuous wave (cw)

dye laser excitation of the ${}^{P}P_{11}(1)$ lines of the $\tilde{A}^2\Pi_{1/2}(0,0,0)-\tilde{X}^2\Sigma^+(0,0,0)$ ($\tilde{\nu}$ =17323.5699 cm$^{-1}$), $\tilde{A}^2\Pi_{1/2}(1,0,0)-\tilde{X}^2\Sigma^+(0,0,0)$ ($\tilde{\nu}$ =17907.9028 cm$^{-1}$) and $[17.73]0.5-\tilde{X}^2\Sigma^+(0,0,0)$ ($\tilde{\nu}$ =17731.9707 cm$^{-1}$) bands. The features marked as "Ex" are emissions occurring at the laser excitation wavelength. The emission near 555.8 nm is due to the $6s^2\ {}^1S_0 \rightarrow 6s6p\ {}^3P_1$ transition of Yb(I) at 17992.007 cm$^{-1}$ which is produced and excited in the laser ablation source.

**Fig. 14.** Proposed re-pumping schemes: a) a simple scheme in which all repumping transitions are driven through the $\tilde{A}^2\Pi_{1/2}(0,0,0)$ state; b) indirect repumping route through the $\tilde{A}^2\Pi_{1/2}(1,0,0)$ state and c) indirect repumping route through the $\tilde{A}^2\Pi_{1/2}(1,0,0)$ and [17.73]0.5 states.

## References

(1) Kozyryev, I.; Hutzler, N. R. Precision Measurement of Time-Reversal Symmetry Violation with Laser-Cooled Polyatomic Molecules. *Phys. Rev. Lett.* **2017**, *119* (13), 133002. https://doi.org/10.1103/PhysRevLett.119.133002.

(2) Maison, D. E.; Skripnikov, L. V.; Flambaum, V. V. Theoretical Study of $^{173}$YbOH to Search for the Nuclear Magnetic Quadrupole Moment. *Phys. Rev. A* **2019**, *100* (3). https://doi.org/10.1103/PhysRevA.100.032514.

(3) Prasannaa, V. S.; Shitara, N.; Sakurai, A.; Abe, M.; Das, B. P. Enhanced Sensitivity of the Electron Electric Dipole Moment from YbOH: The Role of Theory. *Phys. Rev. A* **2019**, *99* (6), 1–6. https://doi.org/10.1103/PhysRevA.99.062502.

(4) Denis, M.; Haase, P. A. B.; Timmermans, R. G. E.; Eliav, E.; Hutzler, N. R.; Borschevsky, A. Enhancement Factor for the Electric Dipole Moment of the Electron in the BaOH and YbOH Molecules. *Phys. Rev. A* **2019**, *99* (4), 42512. https://doi.org/10.1103/physreva.99.042512.

(5) Gaul, K.; Berger, R. Ab Initio Study of Parity and Time-Reversal Violation in Laser-Coolable Triatomic Molecules. *Phys. Rev. A* **2020**, *101* (1), 12508. https://doi.org/10.1103/PhysRevA.101.012508.

(6) Hudson, J. J.; Kara, D. M.; Smallman, I. J.; Sauer, B. E.; Tarbutt, M. R.; Hinds, E. A. Improved Measurement of the Shape of the Electron. *Nature* **2011**, *473* (7348), 493–496. https://doi.org/10.1038/nature10104.

(7) Norrgard, E. B.; Barker, D. S.; Eckel, S.; Fedchak, J. A.; Klimov, N. N.; Scherschligt, J.

Nuclear-Spin Dependent Parity Violation in Optically Trapped Polyatomic Molecules. *Commun. Phys.* **2019**, *2* (1). https://doi.org/10.1038/s42005-019-0181-1.

(8) McCarron, D. J.; Steinecker, M. H.; Zhu, Y.; DeMille, D. Magnetic Trapping of an Ultracold Gas of Polar Molecules. *Phys. Rev. Lett.* **2018**, *121* (1), 13202. https://doi.org/10.1103/physrevlett.121.013202.

(9) Truppe, S.; Williams, H. J.; Hambach, M.; Caldwell, L.; Fitch, N. J.; Hinds, E. A.; Sauer, B. E.; Tarbutt, M. R. Molecules Cooled below the Doppler Limit. *Nat. Phys.* **2017**, *13* (12), 1173–1176. https://doi.org/10.1038/nphys4241.

(10) Lim, J.; Almond, J. R.; Trigatzis, M. A.; Devlin, J. A.; Fitch, N. J.; Sauer, B. E.; Tarbutt, M. R.; Hinds, E. A. Laser Cooled YbF Molecules for Measuring the Electron's Electric Dipole Moment. *Phys. Rev. Lett.* **2018**, *120* (12), 123201. https://doi.org/10.1103/physrevlett.120.123201.

(11) Collopy, A. L.; Ding, S.; Wu, Y.; Finneran, I. A.; Anderegg, L.; Augenbraun, B. L.; Doyle, J. M.; Ye, J. 3D Magneto-Optical Trap of Yttrium Monoxide. *Phys. Rev. Lett.* **2018**, *121* (21), 213201. https://doi.org/10.1103/physrevlett.121.213201.

(12) Melville, T. C.; Coxon, J. A. The Visible Laser Excitation Spectrum of YbOH: The $\tilde{A}\,^2\Pi - \tilde{X}\,^2\Sigma^+$ Transition. *J. Chem. Phys.* **2001**, *115* (15), 6974–6978. https://doi.org/10.1063/1.1404145.

(13) Nakhate, S.; Steimle, T. C.; Pilgram, N. H.; Hutzler, N. R. The Pure Rotational Spectrum of YbOH. *Chem. Phys. Lett.* **2019**, *715*, 105–108. https://doi.org/10.1016/J.CPLETT.2018.11.030.

(14) Steimle, T. C.; Linton, C.; Mengesha, E. T.; Bai, X.; Le, A. T. Field-Free, Stark, and Zeeman Spectroscopy of the Ã 2 Π1/2- X 2 Σ+ Transition of Ytterbium Monohydroxide. *Phys. Rev. A* **2019**, *100* (5), 1–14. https://doi.org/10.1103/PhysRevA.100.052509.

(15) Wang, Hailing;Linton,Colan;Steimle, T. (in preparatin)

(16) Lim, J.; Almond, J. R.; Tarbutt, M. R.; Nguyen, D. T.; Steimle, T. C. The [557]-$X^2\Sigma^+$ and [561]-$X^2\Sigma^+$ Bands of Ytterbium Fluoride, $^{174}$YbF. *J. Mol. Spectrosc.* **2017**, *338*, 81–90. https://doi.org/10.1016/j.jms.2017.06.007.

(17) Maxwell, S. E.; Brahms, N.; deCarvalho, R.; Glenn, D. R.; Helton, J. S.; Nguyen, S. V; Patterson, D.; Petricka, J.; DeMille, D.; Doyle, J. M. High-Flux Beam Source for Cold, Slow Atoms or Molecules. *Phys Rev Lett* **2005**, *95* (17), 173201.

(18) Hutzler, N. R.; Lu, H.-I.; Doyle, J. M. The Buffer Gas Beam: An Intense, Cold, and Slow Source for Atoms and Molecules. *Chem. Rev. (Washington, DC, United States)* **2012**, *112* (9), 4803–4827. https://doi.org/10.1021/cr200362u.

(19) Augenbraun, B. L. .; Lasner, Zack D; Frenet, A. T. .; Sawaoka, H.; Calder, M.; Steimle, Timothy C.; Doyle, J. M. Laser-Cooled Polyatomic Molecules for Improved Electron Electric Dipole Moment Searches. *New J. Phys.* **2019 (accepted)**.

(20) Reilly, N. J.; Schmidt, T. W.; Kable, S. H. Two-Dimensional Fluorescence (Excitation/Emission) Spectroscopy as a Probe of Complex Chemical Environments. *J. Phys. Chem. A* **2006**, *110* (45), 12355–12359. https://doi.org/10.1021/jp064411z.

(21) Kokkin, D. L.; Steimle, T. C.; Demille, D. Branching Ratios and Radiative Lifetimes of the U, L, and i States of Thorium Oxide. *Phys. Rev. A - At. Mol. Opt. Phys.* **2014**, *90* (6).

https://doi.org/10.1103/PhysRevA.90.062503.

(22) Light, J. C.; Carrington Jr., T. Discrete-Variable Representations and Their Utilization. *Adv. Chem. Phys.* **2001**, *114*, 263–310. https://doi.org/10.1002/9780470141731.ch4.

(23) Colbert, D. T.; Miller, W. H. A Novel Discrete Variable Representation for Quantum-Mechanical Reactive Scattering via the S-Matrix Kohn Method. *J. Chem. Phys.* **1992**, *96* (3), 1982–1991. https://doi.org/10.1063/1.462100.

(24) Nooijen, M.; Bartlett, R. J. Equation of Motion Coupled Cluster Method for Electron Attachment. *J. Chem. Phys.* **1995**, *102* (9), 3629–3647. https://doi.org/10.1063/1.468592.

(25) Lu, Q.; Peterson, K. A. Correlation Consistent Basis Sets for Lanthanides: The Atoms La-Lu. *J. Chem. Phys.* **2016**, *145* (5), 054111/1-054111/13. https://doi.org/10.1063/1.4959280.

(26) Dunning Jr., T. H. Gaussian Basis Sets for Use in Correlated Molecular Calculations. I. The Atoms Boron through Neon and Hydrogen. *J. Chem. Phys.* **1989**, *90* (2), 1007–1023. https://doi.org/10.1063/1.456153.

(27) Dyall, K. G. Interfacing Relativistic and Nonrelativistic Methods. IV. One- and Two-Electron Scalar Approximations. *J. Chem. Phys.* **2001**, *115* (20), 9136–9143. https://doi.org/10.1063/1.1413512.

(28) Liu, W.; Peng, D. Exact Two-Component Hamiltonians Revisited. *J. Chem. Phys.* **2009**, *131* (3), 1–5. https://doi.org/10.1063/1.3159445.

(29) Bramley, M. J.; Carrington Jr., T. A General Discrete Variable Methods to Calculate Vibrational Energy Levels of Three- and Four-Atom Molecules. *J. Chem. Phys.* **1993**, *99*

(11), 8519–8541. https://doi.org/10.1063/1.465576.

(30) Changala, B. Density-Functional Thermochemistry. III. The Role of Exact Exchange. *J. Chem. Phys.* **2014**, *140*, 24312. https://doi.org/10.1063/1.4859875.

(31) Nguyen, D. T.; Steimle, T. C.; Kozyryev, I.; Huang, M.; McCoy, A. B. Fluorescence Branching Ratios and Magnetic Tuning of the Visible Spectrum of SrOH. *J. Mol. Spectrosc.* **2018**, *347*, 7–18. https://doi.org/10.1016/j.jms.2018.02.007.

(32) Stanton, J. F.; Gauss, J.; Cheng, L.; Harding, M.; Matthews, D. A.; Szalay, P. G. CFOUR,Coupled-Cluster Techniques for Computational Chemistry.

(33) Stanton, J. F.; Gauss, J. Analytic Energy Gradients for the Equation-of-Motion Coupled-Cluster Method: Implementation and Application to the HCN/HNC System. *J. Chem. Phys* **1994**, *100*, 4695. https://doi.org/10.1063/1.466253.

(34) Stanton, J. F.; Lopreore, C. L.; Gauss, J. The Equilibrium Structure and Fundamental Vibrational Frequencies of Dioxirane. *J. Chem. Phys.* **1998**, *108* (17), 7190–7196. https://doi.org/10.1063/1.476136.

(35) Cheng, L.; Gauss, J. Analytic Second Derivatives for the Spin-Free Exact Two-Component Theory. *J. Chem. Phys.* **2011**, *135* (24). https://doi.org/10.1063/1.3667202.

(36) Presunka, P. I.; Coxon, J. A. Laser Excitation and Dispersed Fluorescence Investigations of the $\tilde{A}^2\Pi - \tilde{X}^2\Sigma^+$ System of SrOH. *Chem. Phys.* **1995**, *190* (1), 97–111. https://doi.org/10.1016/0301-0104(94)00330-D.

(37) Herzberg, G. *Electronic Spectra and Electronic Structure of Polyatomic Molecules (Molecular Spectra and Molecular Structure, Vol. III.*; Van Nostrand, 1966.

(38) Presunka, P. I.; Coxon, J. A. High-Resolution Laser Spectroscopy of Excited Bending Vibrations ($v_2 \leq 2$) of the $\tilde{B}\,^2\Sigma^+$ and $\tilde{X}\,^2\Sigma^+$ Electronic States of SrOH: Analysis of l-Type Doubling and l-Type Resonance. *Can. J. Chem.* **1993**, *71* (10), 1689–1705. https://doi.org/10.1139/v93-211.

(39) Smallman, I. J.; Wang, F.; Steimle, T. C.; Tarbutt, M. R.; Hinds, E. A. Radiative Branching Ratios for Excited States of $^{174}$YbF: Application to Laser Cooling. *J. Mol. Spectrosc.* **2014**, *300*, 3–6. https://doi.org/10.1016/j.jms.2014.02.006.

(40) Bolman, P. S. H.; Brown, J. M. Renner-Teller Effect and Vibronically Induced Bands in the Electronic Spectrum of Isocyanate. *Chem. Phys. Lett.* **1973**, *21* (2), 213–216. https://doi.org/10.1016/0009-2614(73)80121-6.

(41) Gans, B.; Grassi, G.; Merkt, F. Spin-Orbit and Vibronic Coupling in the Ionic Ground State of Iodoacetylene from a Rotationally Resolved Photoelectron Spectrum. *J. Phys. Chem. A* **2013**, *117* (39), 9353–9362. https://doi.org/10.1021/jp310241d.

(42) Smith, T. C.; Li, H.; Hostutler, D. A.; Clouthier, D. J.; Merer, A. J. Orbital Angular Momentum (Renner-Teller) Effects in the $^2\Pi_i$ Ground State of Silicon Methylidyne (SiCH). *J. Chem. Phys.* **2001**, *114* (2), 725–734. https://doi.org/10.1063/1.1331316.

(43) Bunker, P. R.; Jensen, P. *Molecular Symmetry and Spectroscopy, Second Edition*; NRCC, 1998.

(44) Chang, J.-L. A New Formula to Calculate Franck-Condon Factors for Displaced and Distorted Harmonic Oscillators. *J. Mol. Spectrosc.* **2005**, *232* (1), 102–104. https://doi.org/10.1016/j.jms.2005.03.004.

(45) Sharp, T. E.; Rosenstock, H. M. Franck-Condon Factors for Polyatomic Molecules. *J. Chem. Phys.* **1964**, *41* (11), 3453–3463. https://doi.org/10.1063/1.1725748.

(46) Chang, J. L. A New Method to Calculate Franck-Condon Factors of Multidimensional Harmonic Oscillators Including the Duschinsky Effect. *J. Chem. Phys.* **2008**, *128* (17). https://doi.org/10.1063/1.2916717.

(47) Sattasathuchana, T.; Murri, R.; Baldridge, K. K. An Efficient Analytic Approach for Calculation of Multi-Dimensional Franck-Condon Factors and Associated Photoelectron Spectra. *J. Chem. Theory Comput.* **2017**, *13* (5), 2147–2158. https://doi.org/10.1021/acs.jctc.7b00142.

(48) Chen, P. Photoelectron Spectroscopy of Reactive Intermediates; Wiley, 1994; pp 371–425.

(49) Norrgard, E. B.; McCarron, D. J.; Steinecker, M. H.; Tarbutt, M. R.; DeMille, D. Submillikelvin Dipolar Molecules in a Radio-Frequency Magneto-Optical Trap. *Phys. Rev. Lett.* **2016**, *116* (6), 063004/1-063004/6. https://doi.org/10.1103/PhysRevLett.116.063004.

(50) Wilson Jr., E. B.; Decius, J. C.; Cross, P. C. *Molecular Vibrations*; McGraw-Hill Book Co., 1955.

(51) Nguyen, D.-T.; Steimle, T. C.; Kozyryev, I.; Huang, M.; McCoy, A. B. Fluorescence Branching Ratios and Magnetic Tuning of the Visible Spectrum of SrOH. *J. Mol. Spectrosc.* **2018**, *347*, 7–18. https://doi.org/10.1016/j.jms.2018.02.007.

(52) Tandy, J. D.; Wang, J.-G.; Bernath, P. F. High-Resolution Laser Spectroscopy of BaOH

and BaOD: Anomalous Spin-Orbit Coupling in the $\tilde{A}^2\Pi$ State. *J. Mol. Spectrosc.* **2009**, *255* (1), 63–67. https://doi.org/10.1016/j.jms.2009.03.002.

(53) Vamhindi, B. S. D. R.; Nsangou, M. Accurate Ab Initio Potential Energy Curves and Spectroscopic Properties of the Low-Lying Electronic States of $OH^-$ and $SH^-$ Molecular Anions. *Mol. Phys.* **2016**, *114* (14), 2204–2216. https://doi.org/10.1080/00268976.2016.1191690.

(54) Koput, J.; Peterson, K. A. Ab Initio Potential Energy Surface and Vibrational-Rotational Energy Levels of $X^2\Sigma^+$ CaOH. *J. Phys. Chem. A* **2002**, *106* (41), 9595–9599. https://doi.org/10.1021/jp026283u.

**Table 1**

The lifetimes, $\tau_{iv'}$ (ns), branching ratios, $b_{iv',fv''}$ (%), and vibronic transition dipole moments, $\left|\mu_{iv',fv''}\right|$ (D).

| | | $\tilde{X}\,^2\Sigma^+(v_1,v_2^l,v_3)$ | | | | | | | | |
|---|---|---|---|---|---|---|---|---|---|---|
| | | (0,0,0) | (0,1$^1$,0) | (1,0,0) | (0,2$^0$,0) | (1,1$^1$,0) | (2,0,0) | (0,4$^0$,0) | (3,0,0) | (0,6$^0$,0) |
| E($\tilde{X}\,^2\Sigma^+$) cm$^{-1}$ | | 0 | 328 | 529 | 625 | 851 | 1053 | 1140 | 1576 | 1663 |
| $\tilde{A}\,^2\Pi_{1/2}(0,0,0)$ E:17328 cm$^{-1}$ | % | 89.73[a] | | 9.74 | 0.27 | | 0.26 | | <0.05 | |
| τ: 20±1 ns | μ(D) | 5.24 | | 1.81 | 0.30 | | 0.31 | | | |
| [17.33] E:17332 cm$^{-1}$ | % | 89.6 | | 8.8 | 0.6 | | 1.0 | | | |
| τ: 24±2 ns | μ(D) | 4.8 | | 1.6 | 0.4 | | 0.6 | | | |
| [17.637] E:17637 cm$^{-1}$ | % | 74.3 | 10.4 | 6.6 | | 2.7 | 4.2 | 1.8 | | |
| τ: 93±10 ns | μ(D) | | | | | | | | | |
| [17.643] E:17643 cm$^{-1}$ | % | 8.4 | 73.1 | 3.2 | | 15.3 | | | | |
| τ: 35±6 ns | μ(D) | | | | | | | | | |
| [17.68] E: 17681 cm$^{-1}$ | % | 54.1 | 21.8 | 12.8 | | 0.6 | 8.7 | 2.0 | | |
| τ: 89±4 ns | μ(D) | 1.87 | 1.23 | 0.95 | | 0.22 | 0.82 | 0.44 | | |
| [17.73]0.5[b] E:17730 cm$^{-1}$ | % | 58.96 | | 25.18 | 0.60 | | 11.99 | 0.80 | 1.17 | 0.78 |
| τ: 124±4 ns | μ(D) | 1.65 | | 1.13 | 0.18 | | 0.86 | 0.21 | 0.27 | 0.28 |
| [17.90] E:17900 cm$^{-1}$ | % | 8.4 | | 65.1 | 9.4 | | 14.6 | | 2.5 | |
| τ: 26±1 ns | μ(D) | 1.34 | | 3.89 | 1.49 | | 1.92 | | 0.87 | |
| $\tilde{A}\,^2\Pi_{1/2}(1,0,0)$ 17908 cm$^{-1}$ | % | 6.19 | | 64.90 | 11.61 | | 14.98 | | 2.32 | |
| τ: 25±1 ns | μ(D) | 1.17 | | 3.97 | 1.69 | | 1.99 | | 0.82 | |

a) Errors of $b_{iv',fv''}$ values are ± 0.05 for the $\tilde{A}\,^2\Pi_{1/2}(0,0,0)$, $\tilde{A}\,^2\Pi_{1/2}(1,0,0)$ and [17.73]0.5 states and ± 0.1 for all other states.

b) $b_{iv',fv''}$ =0.52 for [17.73]0.5→ $\tilde{X}\,^2\Sigma^+(4,0,0)$ transition

**Table 2.** Energies of the vibrational levels of the $\tilde{X}\,^2\Sigma^+$ state (cm$^{-1}$).

| $\tilde{X}\,^2\Sigma^+(v_1,v_2^l,v_3)$ | Measured[a] | *Ab initio*/DVR[b] |
|---|---|---|
| (0,1$^1$,0) | 329(14) | 322(3) |
| (1,0,0) | 529.33[c](2) | 532(4) |
| (0,2$^0$,0) | 626(7) | 629(0) |
| (0,2$^2$,0) | - | 653(5) |
| (1,1$^1$,0) | 837(-5) | 845(-3) |
| (0,3$^1$,0) | - | 947(-2) |
| (0,3$^3$,0) | - | 996(9) |
| (2,0,0) | 1054(4) | 1059(6) |
| (1,2$^0$,0) | 1140(-9) | 1150(-8) |
| (1,2$^2$,0) | - | 1170(-7) |
| (3,0,0) | 1579(8) | |
| (2,2$^2$,0) | 1658(-11) | |
| (4,0,0) | 2082(-3) | |
| | | |

a) The number in parentheses are the difference between the observed energies and those calculated using least squares optimized values of $\omega_1$=531cm$^{-1}$ , $x_{11}$=-1.9 cm$^{-1}$, $\omega_2$= 310 cm$^{-1}$ and $g_{22}$=5.2 cm$^{-1}$.

b) Ref.[55]. The number in parentheses are the difference between the observed energies and those calculated using least squares optimized values of $\omega_1$=531cm$^{-1}$ , $x_{11}$=-1.9 cm$^{-1}$, $\omega_2$= 310 cm$^{-1}$ and $g_{22}$=5.2 cm$^{-1}$.

c) Ref[14]

**Table 3.** Franck-Condon factors and branching ratios

| Band | Calc. FCFs | | Branching Ratios (%) | |
|---|---|---|---|---|
| | Stretch[a] | Bend[b] | Calc.[c] | Obs. |
| $\tilde{A}^2\Pi_{1/2}(0,0,0)\rightarrow\tilde{X}^2\Sigma^+(0,0,0)$ | 0.8655 | 0.9995 | 87.81 | 89.73 |
| $\tilde{A}^2\Pi_{1/2}(0,0,0)\rightarrow\tilde{X}^2\Sigma^+(1,0,0)$ | 0.1176 | 0.9995 | 10.87 | 9.74 |
| $\tilde{A}^2\Pi_{1/2}(0,0,0)\rightarrow\tilde{X}^2\Sigma^+(0,2^0,0)$ | 0.8655 | 0.0005 | 0.04 | 0.27 |
| $\tilde{A}^2\Pi_{1/2}(0,0,0)\rightarrow\tilde{X}^2\Sigma^+(2,0,0)$ | 0.0150 | 0.9995 | 1.26 | 0.26 |
| $\tilde{A}^2\Pi_{1/2}(0,0,0)\rightarrow\tilde{X}^2\Sigma^+(3,0,0)$ | 0.00016 | 0.9995 | 0.01 | <0.05 |
| $\tilde{A}^2\Pi_{1/2}(1,0,0)\rightarrow\tilde{X}^2\Sigma^+(0,0,0)$ | 0.1301 | 0.9995 | 14.49 | 6.19 |
| $\tilde{A}^2\Pi_{1/2}(1,0,0)\rightarrow\tilde{X}^2\Sigma^+(1,0,0)$ | 0.6339 | 0.9995 | 64.52 | 64.90 |
| $\tilde{A}^2\Pi_{1/2}(1,0,0)\rightarrow\tilde{X}^2\Sigma^+(0,2^0,0)$ | 0.1301 | 0.0005 | <0.05 | 11.61 |
| $\tilde{A}^2\Pi_{1/2}(1,0,0)\rightarrow\tilde{X}^2\Sigma^+(2,0,0)$ | 0.1909 | 0.9995 | 17.73 | 14.98 |
| $\tilde{A}^2\Pi_{1/2}(1,0,0)\rightarrow\tilde{X}^2\Sigma^+(3,0,0)$ | 0.0385 | 0.9995 | 3.25 | 2.32 |
| $\tilde{A}^2\Pi_{1/2}(0,1^1,0)\rightarrow\tilde{X}^2\Sigma^+(0,0,0)$ | 0.8655 | 0 | 0 | 8.4[d] |
| $\tilde{A}^2\Pi_{1/2}(0,1^1,0)\rightarrow\tilde{X}^2\Sigma^+(0,1^1,0)$ | 0.8655 | 0.9984 | 88.8 | 73.1 |
| $\tilde{A}^2\Pi_{1/2}(0,1^1,0)\rightarrow\tilde{X}^2\Sigma^+(1,0,0)$ | 0.1176 | 0 | 0 | 3.2 |
| $\tilde{A}^2\Pi_{1/2}(0,1^1,0)\rightarrow\tilde{X}^2\Sigma^+(1,1^1,0)$ | 0.8655 | 0.0015 | 11.3 | 15.3 |
| $\tilde{A}^2\Pi_{1/2}(0,1^0,0)\rightarrow\tilde{X}^2\Sigma^+(2,0,0)$ | 0.0150 | 0 | 0 | < 1% |
| | | | | |

a) Two dimensional $\left|\left\langle\tilde{A}^2\Pi,v_1,v_3\middle|\tilde{X}^2\Sigma^+,v_1,v_3\right\rangle\right|^2$ in the harmonic approximation.

b) One dimensional $\left|\left\langle\tilde{A}^2\Pi,v_2\middle|\tilde{X}^2\Sigma^+,v_2\right\rangle\right|^2$ in the harmonic approximation.

c) Eq. 7.

d) The observed data for the [17.643] level.

Figure 1

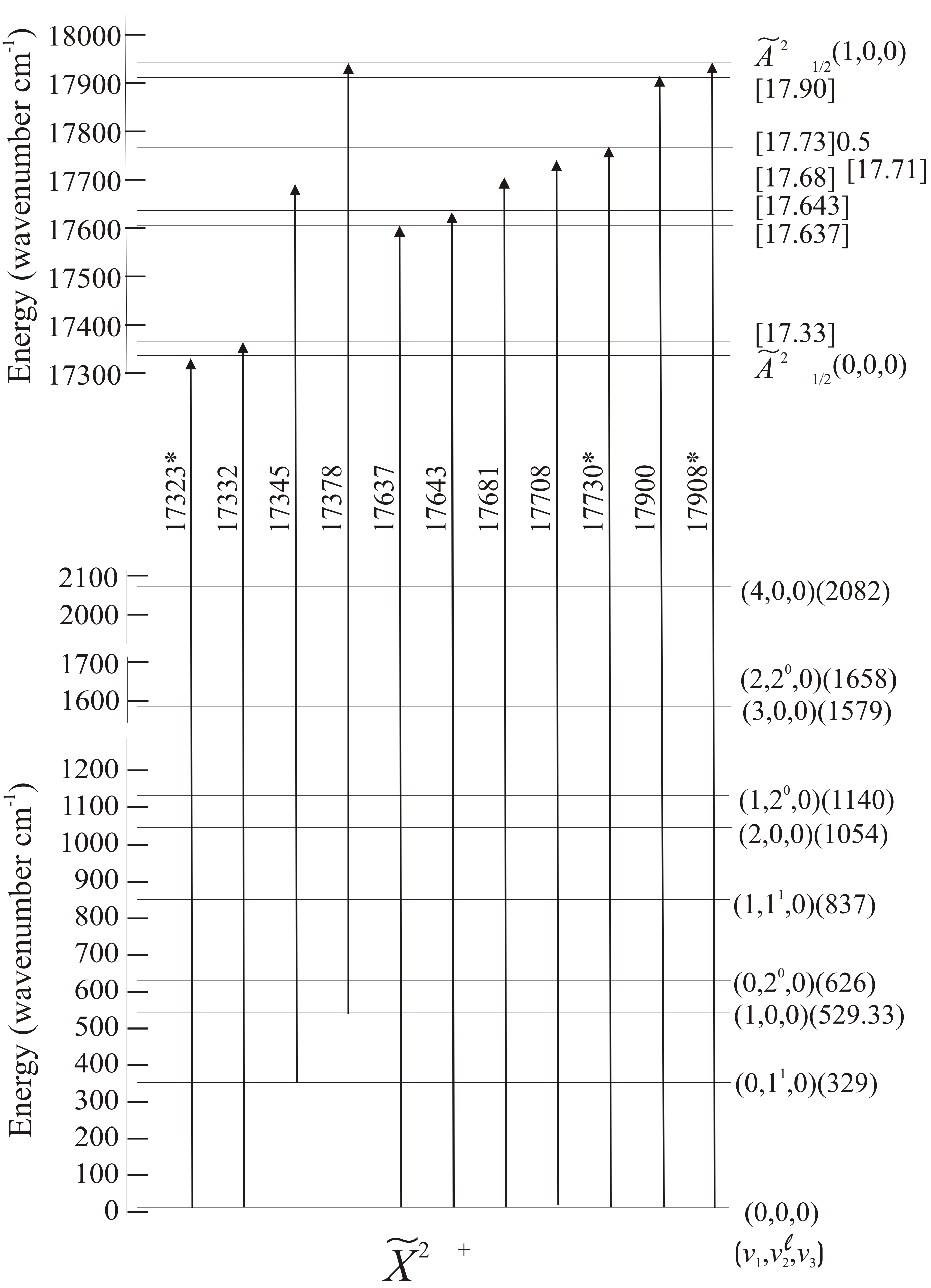

# Figure 2

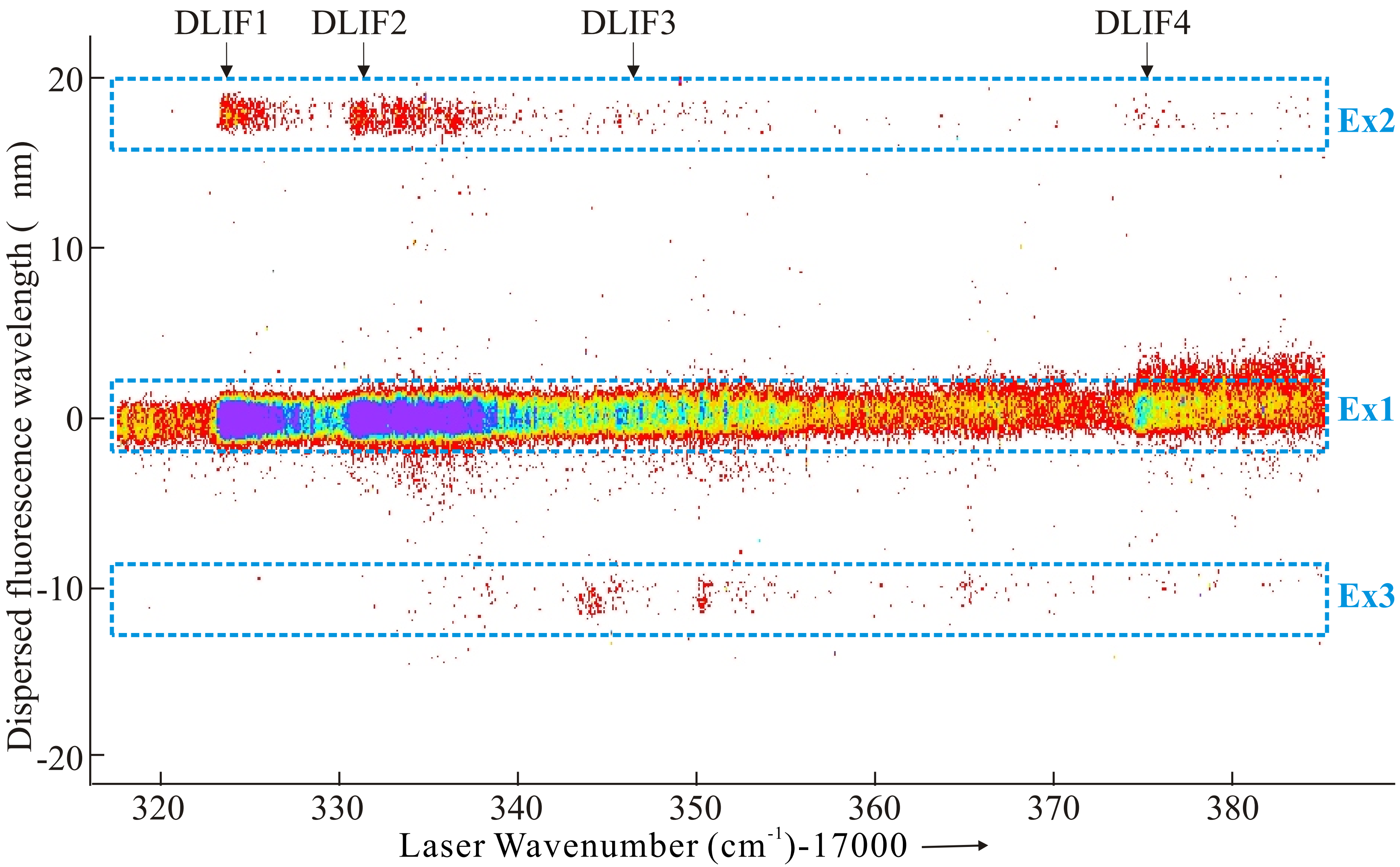

Figure 3

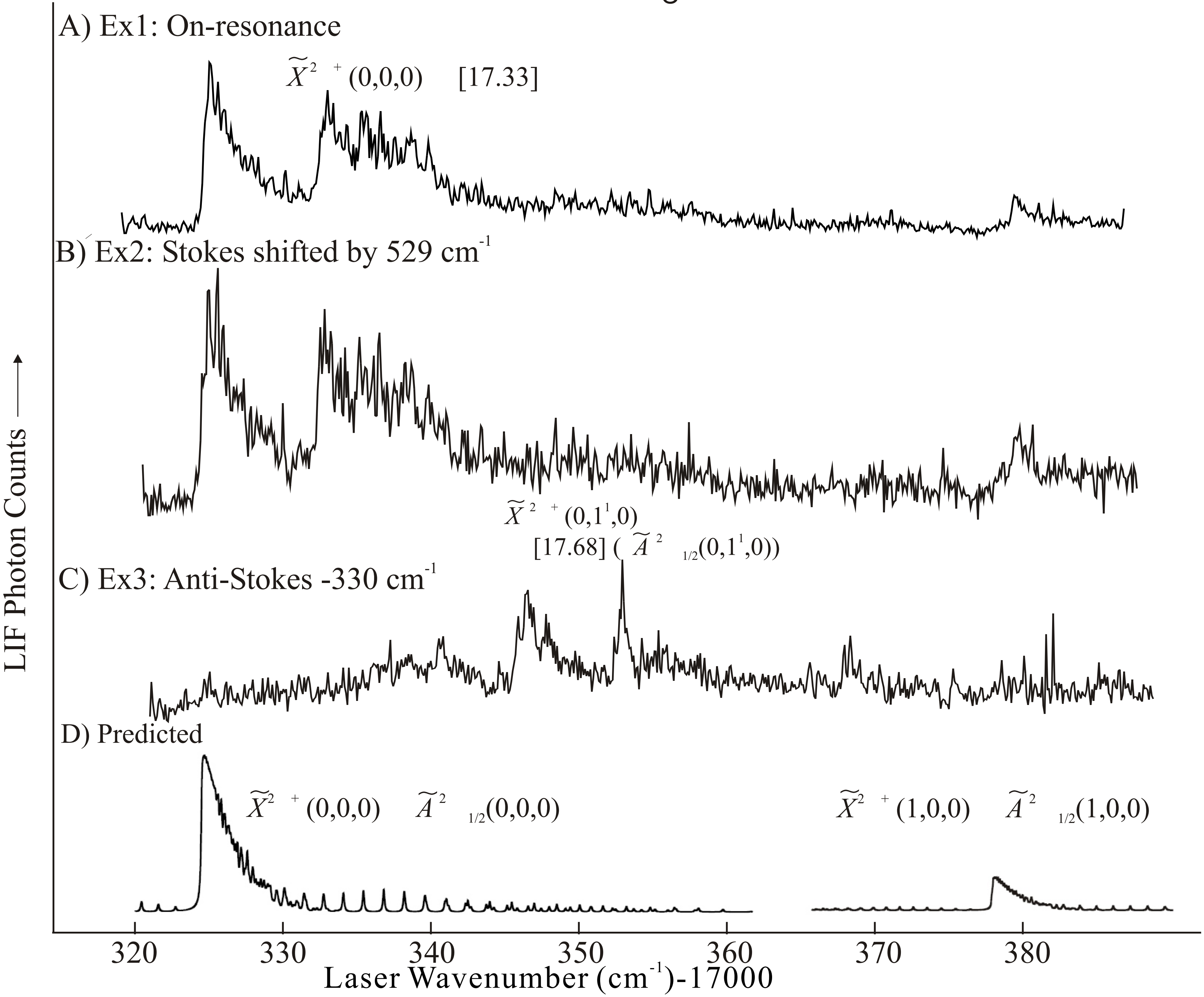

A) Ex1: On-resonance
X̃ 2 + (0,0,0) [17.33]
B) Ex2: Stokes shifted by 529 cm-1
X̃ 2 + (0,1 1,0)
[17.68] ( Ã 2 1/2(0,1 1,0))
C) Ex3: Anti-Stokes -330 cm-1
D) Predicted
X̃2 + (0,0,0) Ã2 1/2(0,0,0)
X̃2 + (1,0,0) Ã2 1/2(1,0,0)
LIF Photon Counts
320
330
340
350
360
370
380
Laser Wavenumber (cm-1)-17000

Figure 4

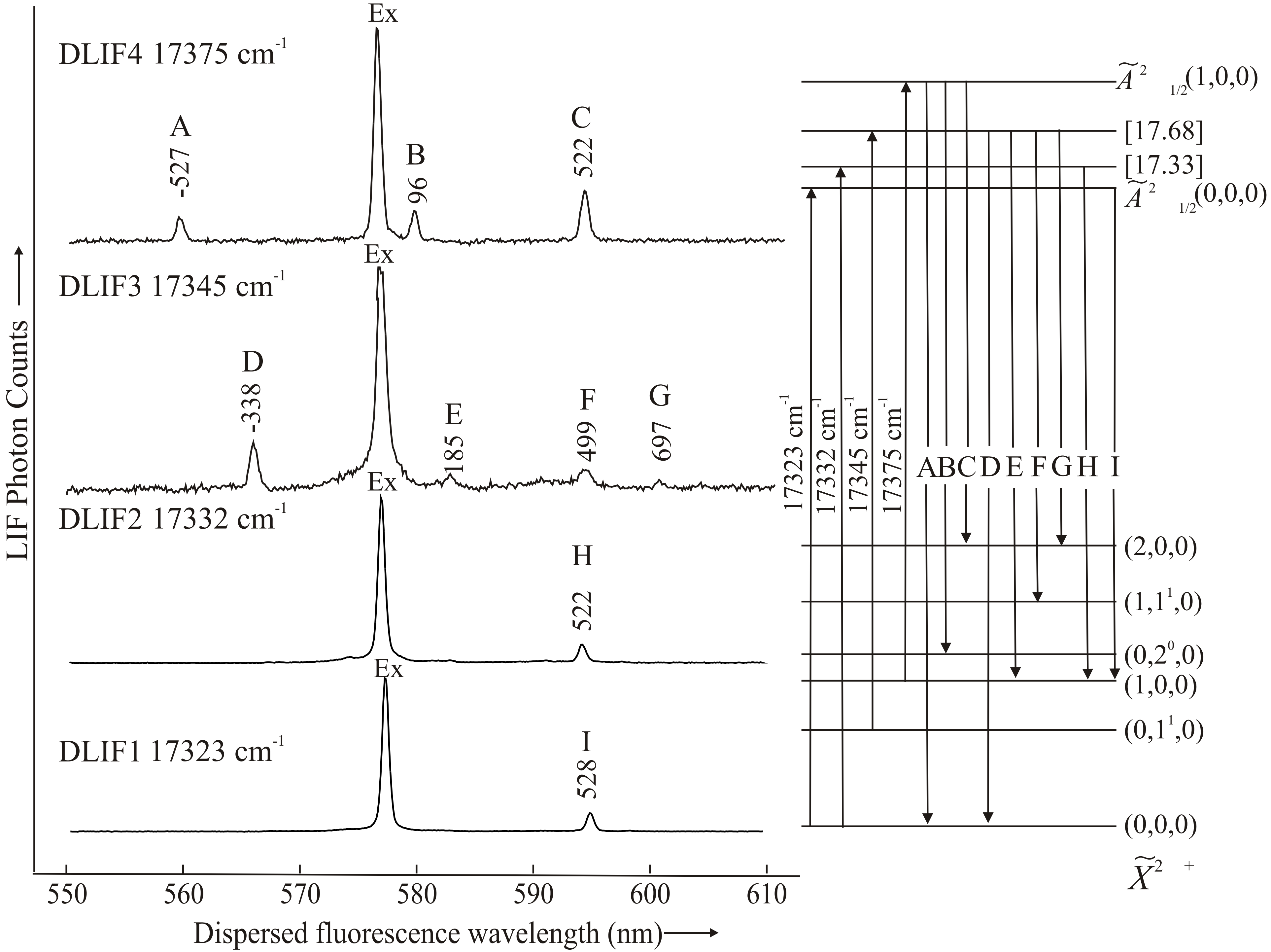

Figure 5

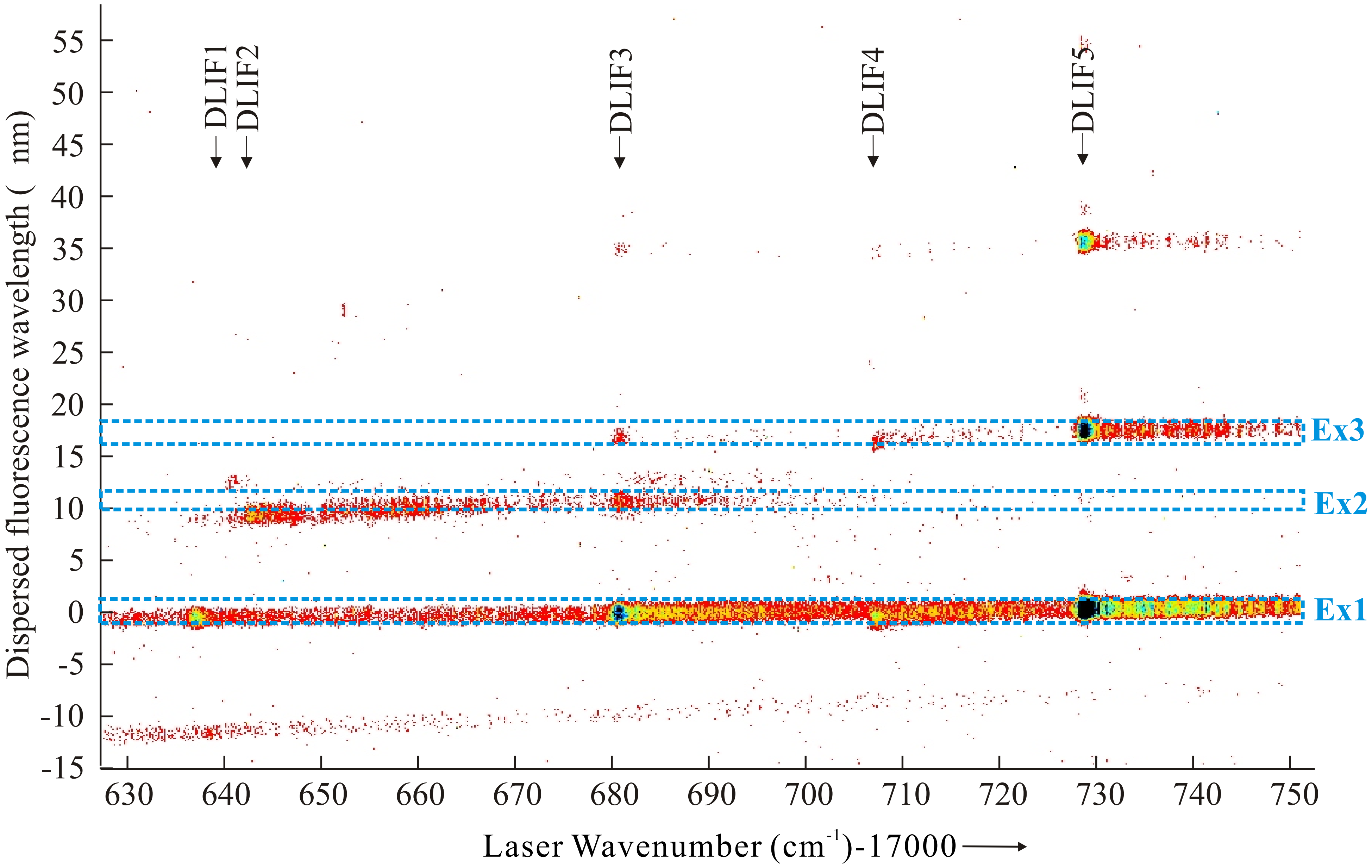

Figure 6

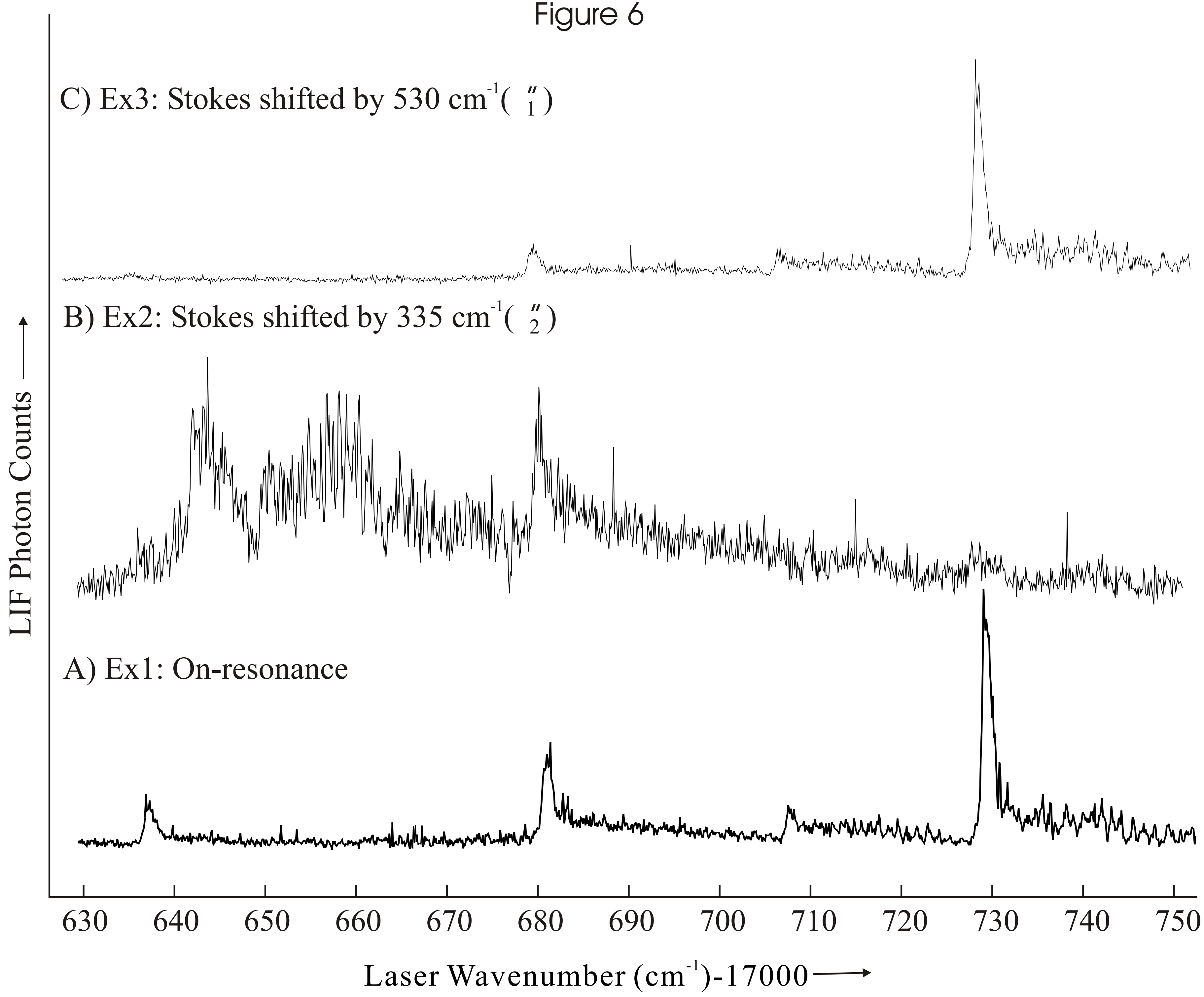

C) Ex3: Stokes shifted by 530 cm$^{-1}$( $''_1$ )
B) Ex2: Stokes shifted by 335 cm$^{-1}$( $''_2$ )
A) Ex1: On-resonance
LIF Photon Counts
630
640
650
660
670
680
690
700
710
720
730
740
750
Laser Wavenumber (cm$^{-1}$)-17000

Figure 7

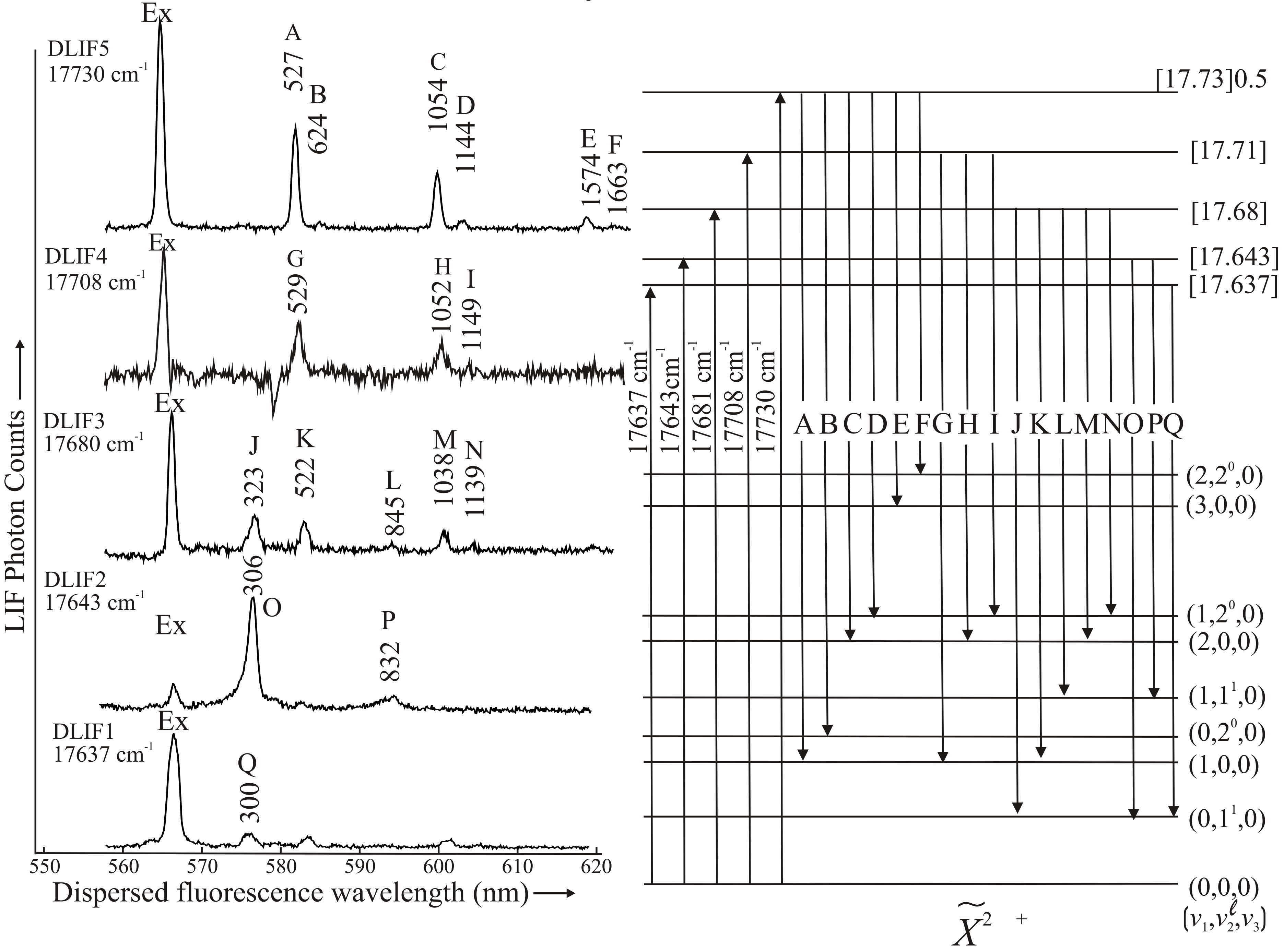

DLIF5
17730 cm-1
Ex
A
527
B
624
C
1054
D
1144
E
1574
F
1663
DLIF4
17708 cm-1
Ex
G
529
H
1052
I
1149
DLIF3
17680 cm-1
Ex
J
323
K
522
L
845
M
1038
N
1139
DLIF2
17643 cm-1
Ex
306
O
P
832
DLIF1
17637 cm-1
Ex
Q
300
LIF Photon Counts
550 560 570 580 590 600 610 620
Dispersed fluorescence wavelength (nm)
[17.73]0.5
[17.71]
[17.68]
[17.643]
[17.637]
17637 cm-1
17643cm-1
17681 cm-1
17708 cm-1
17730 cm-1
A B C D E F G H I J K L M N O P Q
(2,2⁰,0)
(3,0,0)
(1,2⁰,0)
(2,0,0)
(1,1¹,0)
(0,2⁰,0)
(1,0,0)
(0,1¹,0)
(0,0,0)
(v1,v2ℓ,v3)
X̃2 +

# Figure 8

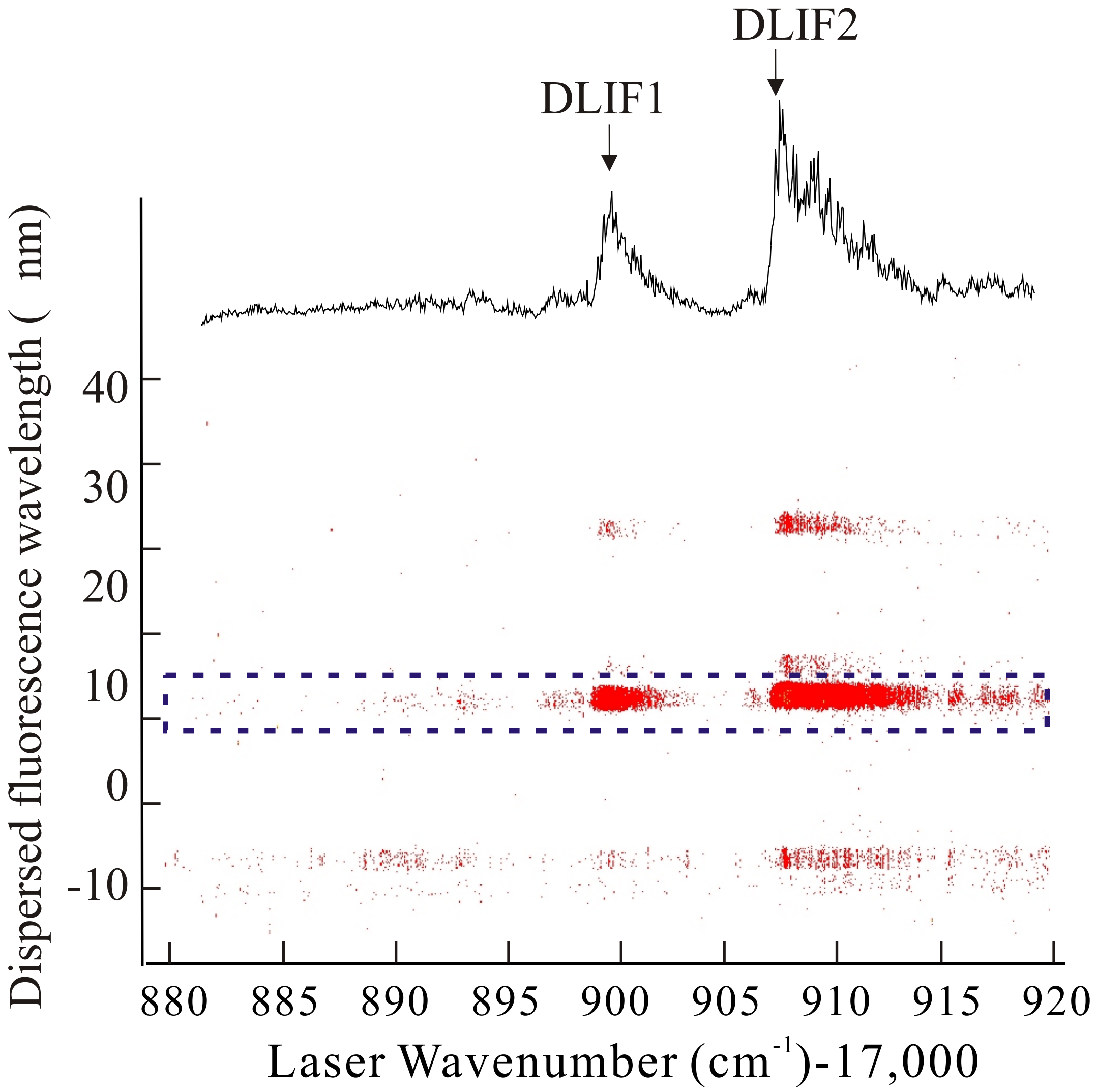

# Figure 9

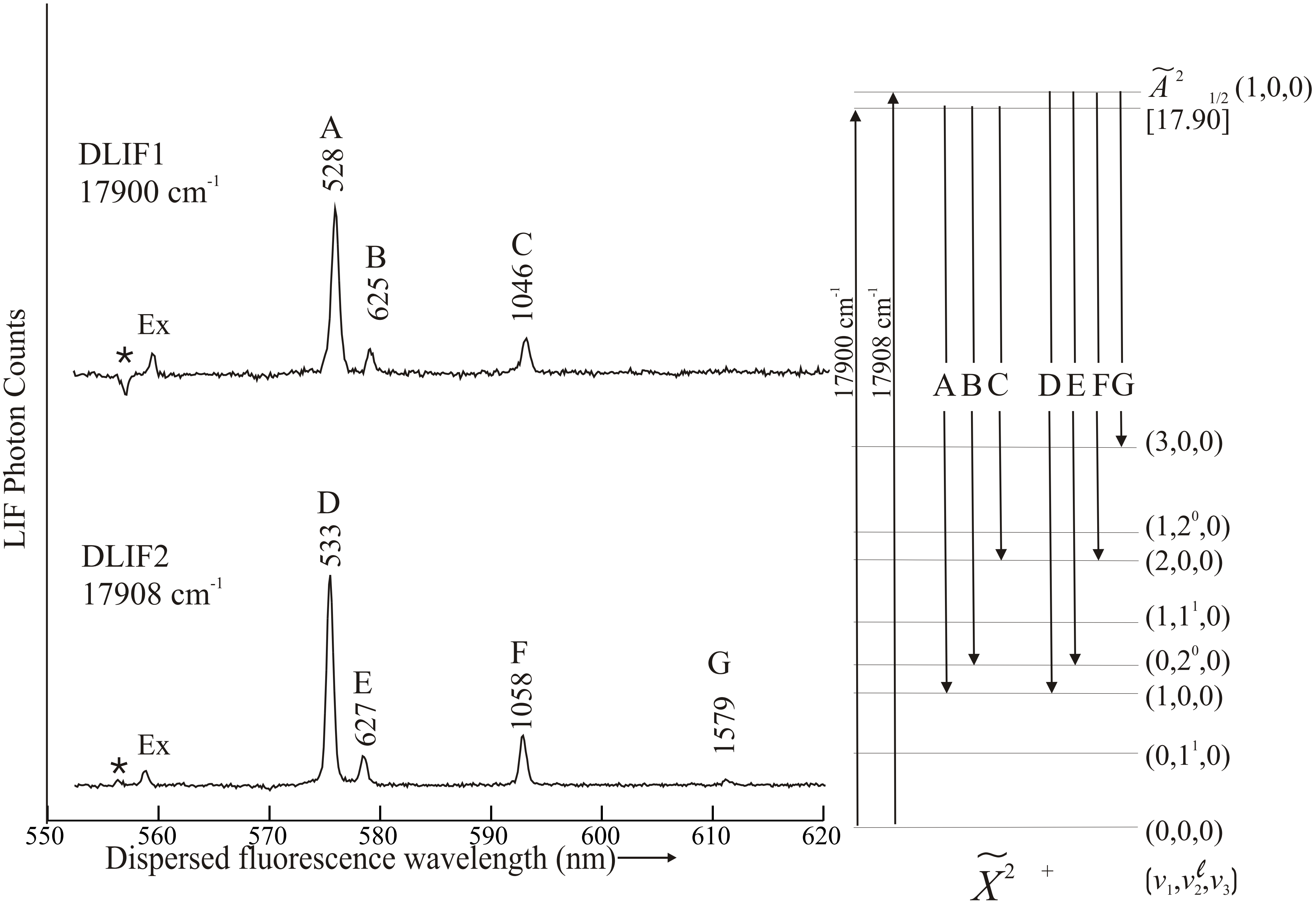

# Figure 10

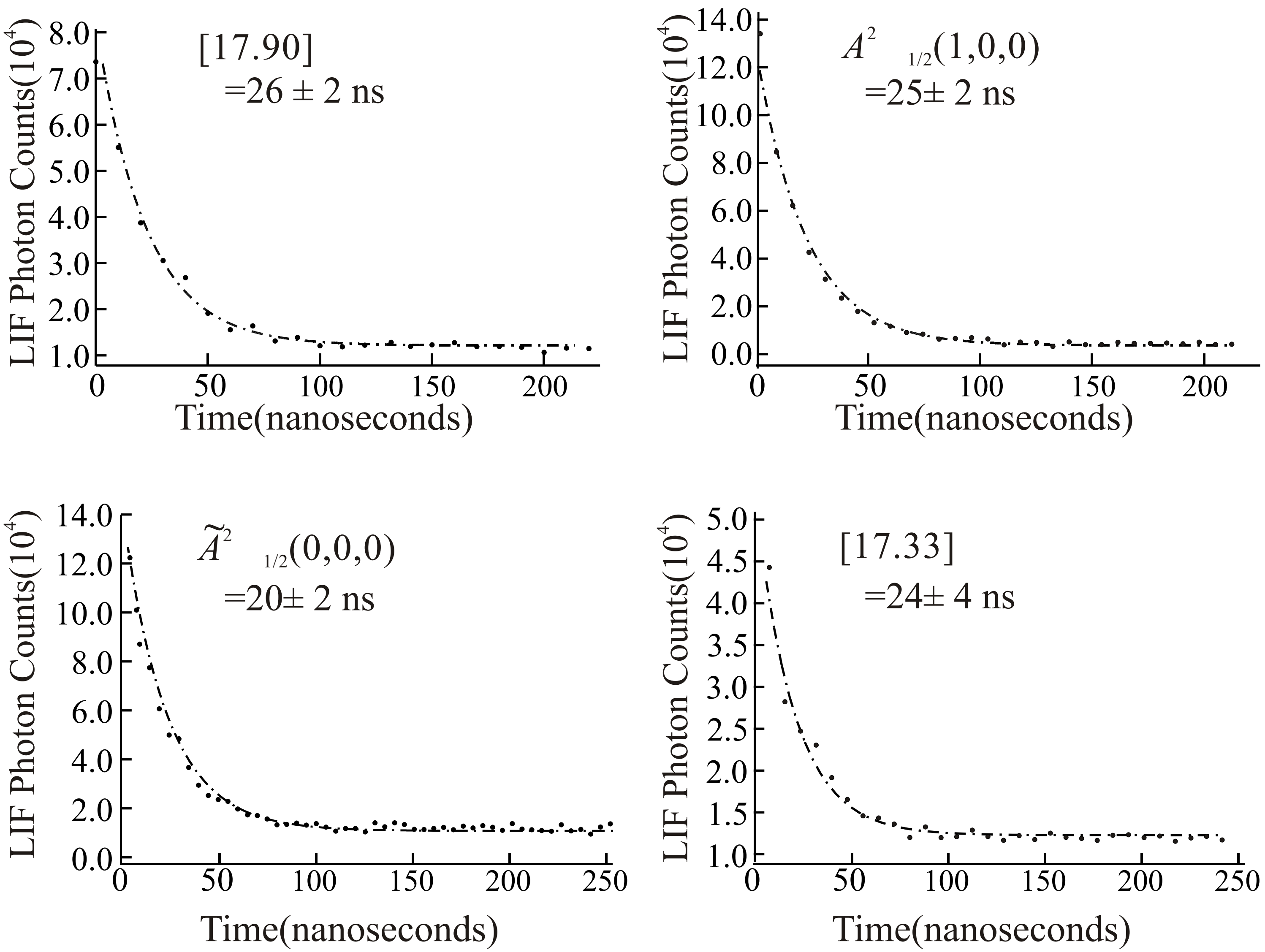

Figure 11

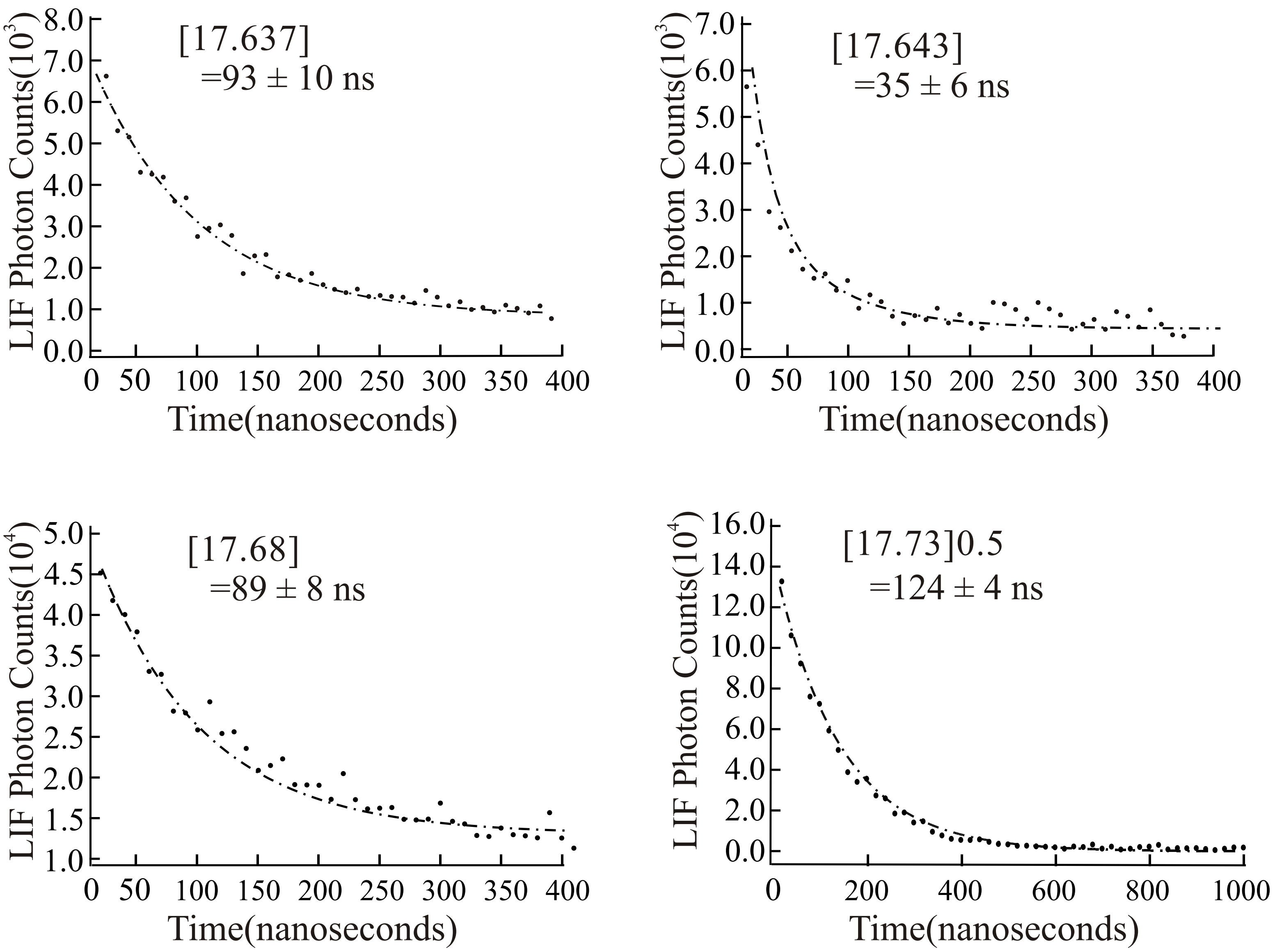
[17.637]
=93 ± 10 ns
LIF Photon Counts($10^3$)
Time(nanoseconds)
[17.643]
=35 ± 6 ns
LIF Photon Counts($10^3$)
Time(nanoseconds)
[17.68]
=89 ± 8 ns
LIF Photon Counts($10^4$)
Time(nanoseconds)
[17.73]0.5
=124 ± 4 ns
LIF Photon Counts($10^4$)
Time(nanoseconds)

Figure 12

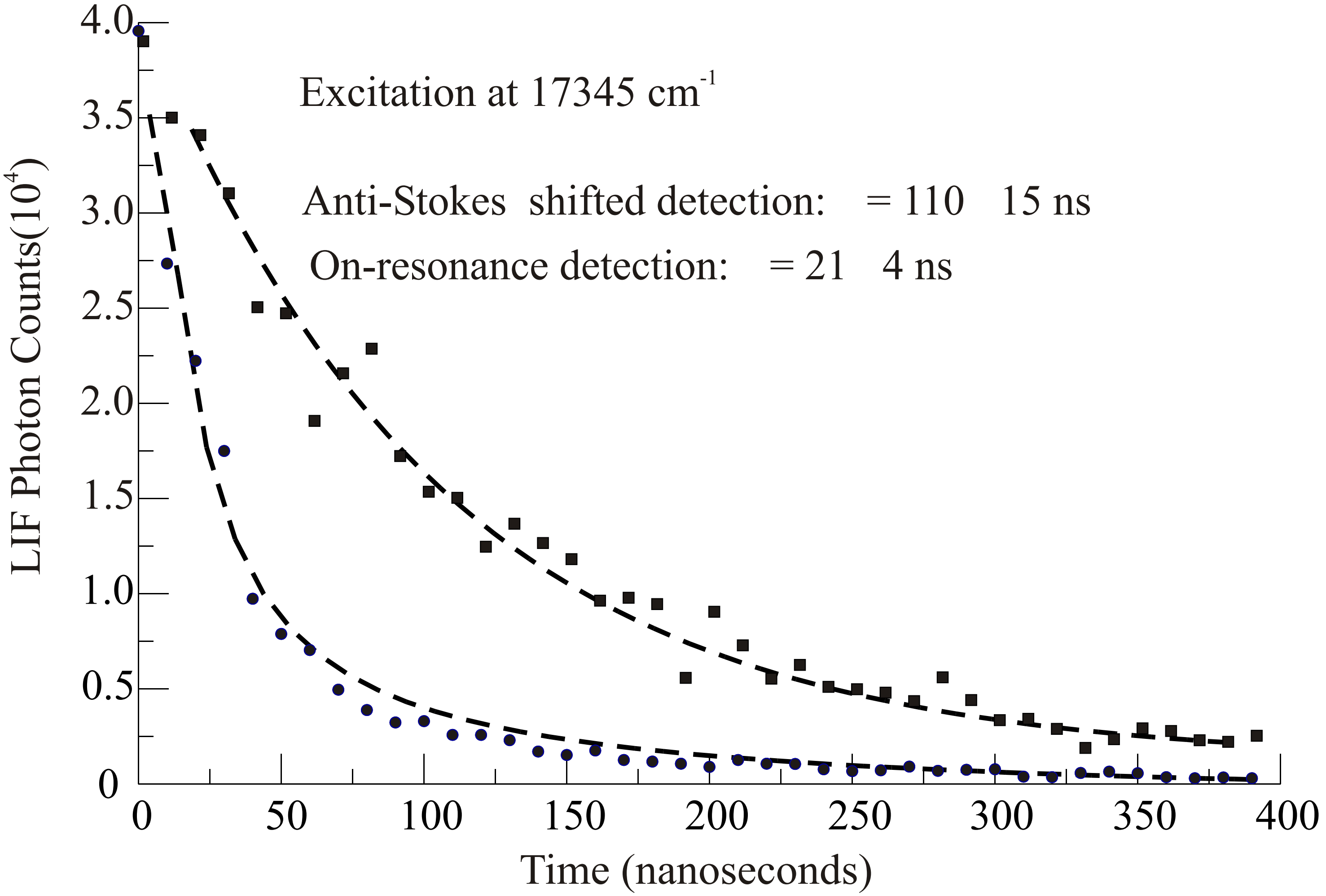

Figure 13

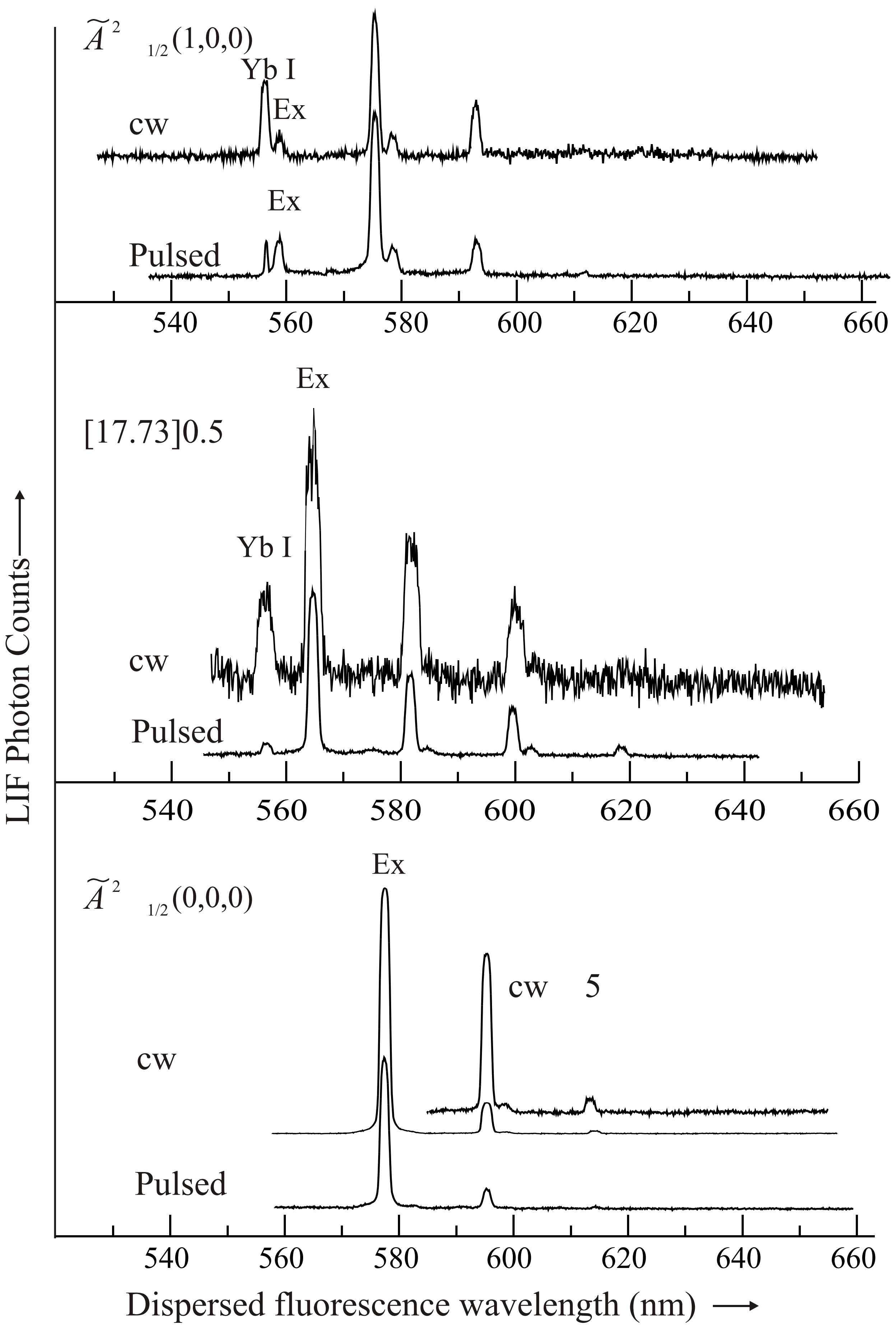

# Figure 14

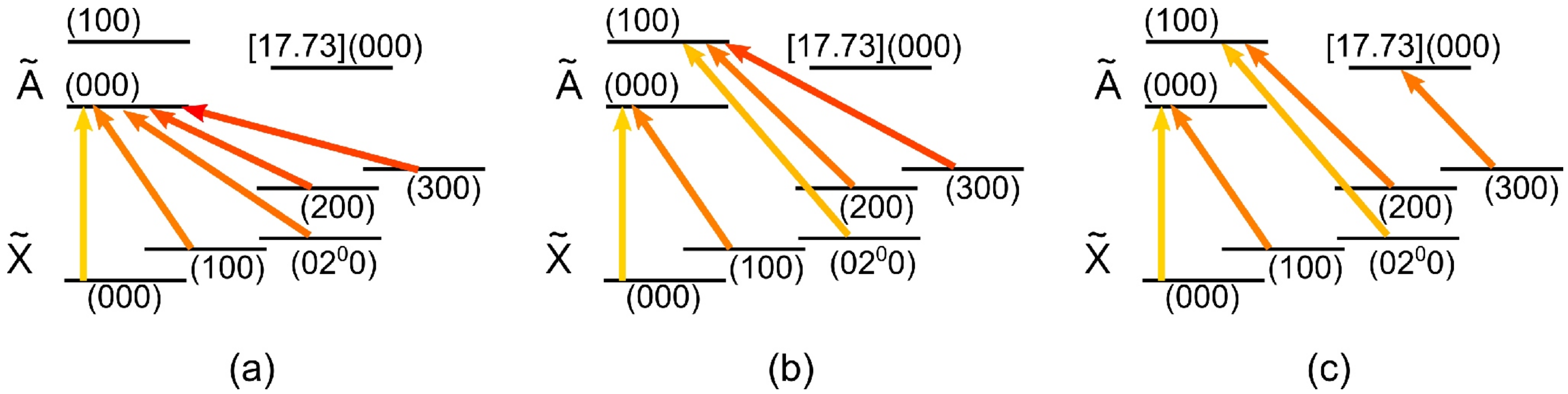

## Supporting Information

The discrete variable representation calculations reported here employed a standard formulation for rectilinear coordinates. In the present calculations, we only included vibrations and did not consider the Coriolis term in the molecular Hamiltonian. As the rotational constant of YbOH is small (around 0.24 $cm^{-1}$), the approximation of not including the Coriolis term is expected to have a minor effect on the computed vibrational energy levels. The harmonic vibration frequencies and dimensionless reduced normal coordinates (in Bohr) as well as equilibrium structure in Cartesian coordinate (in Bohr) of YbOH are shown in Table S1.

**Table S1**

| | | | | |
|---|---|---|---|---|
| **Bending** 340.27 $cm^{-1}$ | **Yb** | **0.0014963090** | **0.0000000000** | **0.0000000000** |
| | **O** | **-0.0512939653** | **0.0000000000** | **0.0000000000** |
| | **H** | **0.5558269458** | **0.0000000000** | **0.0000000000** |
| **Bending** 340.27 $cm^{-1}$ | **Yb** | **0.0000000000** | **-0.0014963090** | **0.0000000000** |
| | **O** | **0.0000000000** | **0.0512939653** | **0.0000000000** |
| | **H** | **0.0000000000** | **-0.5558269458** | **0.0000000000** |
| **Yb-O Stretching** 548.35 $cm^{-1}$ | **Yb** | **0.0000000000** | **0.0000000000** | **0.0106020780** |
| | **O** | **0.0000000000** | **0.0000000000** | **-0.1083813177** |
| | **H** | **0.0000000000** | **0.0000000000** | **-0.1097049772** |
| **O-H Stretching** 4016.83 $cm^{-1}$ | **Yb** | **0.0000000000** | **0.0000000000** | **0.0000107746** |
| | **O** | **0.0000000000** | **0.0000000000** | **-0.0106495013** |
| | **H** | **0.0000000000** | **0.0000000000** | **0.1671557484** |
| **Equilibrium Structure** | **Yb** | **0.0000000000** | **0.0000000000** | **-0.3536542927** |
| | **O** | **0.0000000000** | **0.0000000000** | **3.5114359323** |
| | **H** | **0.0000000000** | **0.0000000000** | **5.3075742317** |

*Ab initio* calculations were carried out for 2187 grid points in a rectangular grid (11 evenly spaced points covering the range [3.46509 bohr, 4.46509 bohr] for Yb-O bond length, 11 points covering the range [1.59614 bohr, 2.09614 bohr] for O-H bond length, and 18 points covering the range [95.0°, 180.0°] for the Yb-O-H angle). An analytic potential function was obtained by fitting these *ab initio* energies into a six-order polynomial in terms of three internal coordinate displacements $r_1 - r_{1e}$, $r_2 - r_{2e}$ and $\theta - \theta_e$, where $r_1$ and $r_2$ refer to Yb-O and O-H bond lengths, respectively, i.e., $f(r_1, r_2, \theta) = \sum_{i,j,k}(1/i!\,j!\,k!)A_{ijk}(r_1 - r_{1e})^i(r_2 - r_{2e})^j(\theta - \theta_e)^k$. The equilibrium structure parameters $r_{1e}$, $r_{2e}$ and $\theta_e$ are shown in Table S2.

**Table S2 Geometrical parameters for the equilibrium structure of YbOH**

| | $r_{1e}$/bohr | $r_{2e}$/bohr | $\theta_e$/degree |
|---|---|---|---|
| **Structure Constants** | **3.86509** | **1.79614** | **180.00** |

The values of the coefficients $A_{ijk}$ are shown in Table S3. The energy unit is $cm^{-1}$.

**Table S3 Fitting parameters for the potential energy function of the electronic ground state of YbOH**

| $i$ | $j$ | $k$ | $A_{ijk}$ | $i$ | $j$ | $k$ | $A_{ijk}$ |
|---|---|---|---|---|---|---|---|
| 0 | 0 | 0 | -3101899567 | 2 | 0 | 0 | 38704.07428 |
| 0 | 0 | 1 | -0.383898023 | 2 | 0 | 1 | 7.105794159 |
| 0 | 0 | 2 | 0.77924781 | 2 | 0 | 2 | 3.224491406 |
| 0 | 0 | 3 | -0.001854314 | 2 | 0 | 3 | 0.132594904 |
| 0 | 0 | 4 | 0.000873252 | 2 | 0 | 4 | 0.002441251 |
| 0 | 0 | 5 | 0.00012016 | 2 | 1 | 0 | -3458.943316 |
| 0 | 0 | 6 | 5.5125E-06 | 2 | 1 | 1 | 11.23176778 |
| 0 | 1 | 0 | -1.430906661 | 2 | 1 | 2 | 1.979279247 |
| 0 | 1 | 1 | -1.424442404 | 2 | 1 | 3 | 0.050370627 |
| 0 | 1 | 2 | -1.218293541 | 2 | 2 | 0 | -2805.753496 |
| 0 | 1 | 3 | -0.04261763 | 2 | 2 | 1 | -41.48778013 |
| 0 | 1 | 4 | -0.003211632 | 2 | 2 | 2 | -1.016857367 |
| 0 | 1 | 5 | -8.65572E-05 | 2 | 3 | 0 | 2446.183784 |
| 0 | 2 | 0 | 127025.4725 | 2 | 3 | 1 | 30.45977615 |
| 0 | 2 | 1 | -1.698845837 | 2 | 4 | 0 | -8770.407632 |
| 0 | 2 | 2 | -0.442797787 | 3 | 0 | 0 | -89060.28281 |
| 0 | 2 | 3 | -0.049911514 | 3 | 0 | 1 | 31.16345449 |
| 0 | 2 | 4 | -0.002422788 | 3 | 0 | 2 | 1.808260952 |
| 0 | 3 | 0 | -464085.3901 | 3 | 0 | 3 | 0.025845576 |
| 0 | 3 | 1 | -8.238106213 | 3 | 1 | 0 | 8305.851882 |
| 0 | 3 | 2 | -1.528665475 | 3 | 1 | 1 | 73.86093166 |
| 0 | 3 | 3 | -0.062052299 | 3 | 1 | 2 | 0.922400013 |
| 0 | 4 | 0 | 1545851.083 | 3 | 2 | 0 | 6563.620737 |
| 0 | 4 | 1 | 15.91409111 | 3 | 2 | 1 | -35.87229531 |
| 0 | 4 | 2 | 1.851241631 | 3 | 3 | 0 | 2833.934456 |
| 0 | 5 | 0 | -5485880.256 | 4 | 0 | 0 | 171368.8716 |
| 0 | 5 | 1 | -128.9109633 | 4 | 0 | 1 | 48.72867329 |
| 0 | 6 | 0 | 16484825.49 | 4 | 0 | 2 | -1.529885755 |
| 1 | 0 | 0 | -1.734526417 | 4 | 1 | 0 | -18101.95738 |
| 1 | 0 | 1 | 0.498466484 | 4 | 1 | 1 | -28.05395191 |
| 1 | 0 | 2 | -2.339761015 | 4 | 2 | 0 | -21342.18234 |
| 1 | 0 | 3 | 0.067250889 | 5 | 0 | 0 | -284703.1294 |
| 1 | 0 | 4 | 0.008799531 | 5 | 0 | 1 | -259.9039203 |
| 1 | 0 | 5 | 0.000247944 | 5 | 1 | 0 | 39244.10432 |
| 1 | 1 | 0 | 886.0437053 | 6 | 0 | 0 | 302191.8983 |
| 1 | 1 | 1 | -6.861971441 | | | | |
| 1 | 1 | 2 | -0.971948193 | | | | |
| 1 | 1 | 3 | -0.040795813 | | | | |
| 1 | 1 | 4 | -0.000331089 | | | | |
| 1 | 2 | 0 | 220.8883148 | | | | |
| 1 | 2 | 1 | -7.395075627 | | | | |
| 1 | 2 | 2 | -1.955444977 | | | | |
| 1 | 2 | 3 | -0.065314243 | | | | |
| 1 | 3 | 0 | -3145.328395 | | | | |
| 1 | 3 | 1 | -25.69432933 | | | | |
| 1 | 3 | 2 | -0.468191019 | | | | |
| 1 | 4 | 0 | 5862.787512 | | | | |
| 1 | 4 | 1 | -23.92446116 | | | | |
| 1 | 5 | 0 | 2945.482351 | | | | |